\documentclass[aps,pra, twocolumn, groupedaddress, twoside]{revtex4-2}

\usepackage{graphicx}
\usepackage{amsfonts}
\usepackage{amsmath}
\usepackage{amssymb}

\usepackage{xcolor}
\definecolor{grey}{rgb}{0.7,0.7,0.7}
\definecolor{db}{rgb}{0,0,0.5}

\def\bs{\boldsymbol}
\def\bse{\begin{subequations}}
\def\ese{\end{subequations}}

\usepackage{fancyhdr}
 
\begin{document}



\title{\colorbox{db}{\parbox{\linewidth}{  \centering \parbox{0.9\linewidth}{ \textbf{\centering\Large{\color{white}{ \vskip0.2em{Optical Singularity Dynamics and Spin-Orbit Interaction due to a Normal-Incident Optical Beam Reflected at a Plane Dielectric Interface}\vskip0.8em}}}}}}}




\author{Anirban Debnath}
\email[]{anirban.debnath090@gmail.com}
\affiliation{School of Physics, University of Hyderabad, Hyderabad 500046, India}

\author{Nitish Kumar}
\thanks{These authors have contributed equally to this paper.}
\affiliation{School of Physics, University of Hyderabad, Hyderabad 500046, India}

\author{Upasana Baishya}
\thanks{These authors have contributed equally to this paper.}
\affiliation{School of Physics, University of Hyderabad, Hyderabad 500046, India}

\author{Nirmal K. Viswanathan}
\email[]{nirmalsp@uohyd.ac.in}
\affiliation{School of Physics, University of Hyderabad, Hyderabad 500046, India}


\date{\today}

\begin{abstract}
\noindent{\color{grey}{\rule{0.784\textwidth}{1pt}}}
\vspace{-0.8em}

The degenerate case of normal incidence and reflection of an optical beam (both paraxial and non-paraxial) at a plane isotropic dielectric interface, which is azimuthally symmetric in terms of the momentum-spatial variation of Fresnel coefficients but not in terms of the fundamental polarization inhomogeneity of the incident field, requires in-depth analyses. In this paper, we use the reflection and transmission coefficient matrix formalism to derive an exact field expression of a normal-reflected diverging beam. The availability of the exact field information allows controlled variations of the system parameters, leading to significant dynamics of phase and polarization singularities hitherto unanticipated in the literature. 
We carry out a detailed exploration of these dynamics in our simulated system, and also verify them experimentally by using an appropriate setup.
We then use Barnett's formalism to determine the associated orbital angular momentum (OAM) fluxes, leading to a subtle interpretation and mathematical characterization of spin-orbit interaction (SOI) in the system. 
Our work thus represents a non-trivial unification of the most fundamental electromagnetic reflection/transmission problem at a plane dielectric interface and 
the emerging areas of optical singularity dynamics with their understanding in terms of OAM flux and SOI.
The normal-incidence--retro-reflection geometry being especially amenable to applications, these beam-field phenomena are anticipated to have applications in interface characterization, particle rotation/manipulation and other nano-optical processes.
{\color{grey}{\rule{0.784\textwidth}{1pt}}}
\end{abstract}


\maketitle



\tableofcontents

{\color{grey}{\noindent\rule{\linewidth}{1pt}}}


\section{Introduction} \label{Sec_Intro}

Any real optical beam, unlike an ideal plane wave, is mathematically composed of infinite constituent ideal plane waves \cite{SalehTeich}. These plane waves are associated to their individual wavevectors, all of which together create a non-point-like distribution in the momentum space. In a reflection-transmission analysis of such a `composite' optical beam --- even at the simplest of surfaces such as a plane isotropic dielectric interface --- each constituent plane wave must be transformed and subsequently combined appropriately to obtain the correct forms of the complete reflected and transmitted beam fields. The spatial dispersion of Fresnel coefficients \cite{BARev}, or equivalently, the momentum-spatial variation of the reflection and transmission coefficient matrices \cite{ADNKVrt2020} come into play in this context, giving rise to distorted complicated reflected and transmitted field profiles. As a result, the centroids of the reflected and transmitted intensity distributions are generally shifted from the geometrically expected positions. 
The longitudinal shift, which is along the central plane of incidence, is known as the Goos-H\"anchen (GH) shift; and the transverse shift, which is perpendicular to the central plane of incidence, is known as the Imbert-Fedorov (IF) shift \cite{GH, Artmann, RaJW, AntarYM, McGuirk, ChanCC, Porras, AielloArXiv, Fedorov, Schilling, Imbert, Player, FVG, Liberman, Onoda, Bliokh2006, Bliokh2007, HostenKwiat, AielloArXiv2, Aiello2008, Merano2009, Aiello2009, Qin2011, BARev, GotteLofflerDennis, XieSHELinIF, ADNKVrt2020}.

A significant degenerate situation, however, is observed when the central incident wavevector is normal to the dielectric interface (referred to as the normal incidence/reflection case in the present paper). Due to normal incidence of the central wavevector, no specific central plane of incidence is obtained, with respect to which the longitudinal and transverse directions would be defined. Physically, the conventional GH and IF shifts disappear in this situation \cite{BARev}. However, the azimuthally symmetric variation of Fresnel coefficients (with respect to the orientations of the constituent wavevectors), combined with the inhomogeneous polarization profile of the composite incident beam field, gives rise to substantially complex beam-field phenomena that accompany the complicated reflected and transmitted field profiles.

Standard analyses of the normal incidence case are present in the literature \cite{NovotnyFocusImage2001, PetrovFocusedBeamRT2005, NanoOpticsBook, WangNormalShift2021, VortexBrewster}. 
However, to our knowledge, a detailed exploration of phase and polarization singularity dynamics in a normal-reflected beam field has not been worked out yet. Understanding the nature of these fundamental electromagnetic-optical phenomena would offer novel methodologies to explore interface characteristics and emerging nano-optical processes such as particle trapping and rotation/manipulation.
Additionally, one may anticipate that the exact information on orbital angular momentum (OAM) flux would aid to such nano-optical applications, because the OAM flux density rather than the OAM density is responsible for OAM transfer across any interface \cite{Barnett2002, Gbur}. 
The OAM flux density is related to the OAM density via a continuity equation; and the inconsistencies posed by OAM density in the decomposition of total angular momentum (AM) into spin and orbital parts is resolved by OAM flux density.
However, a complete characterization of OAM flux in a normal-reflected beam field is not currently present in the literature.

In the present paper, we aim to unravel the above-mentioned unexplored areas by analyzing normal optical beam reflection at a plane isotropic dielectric interface by using the reflection and transmission coefficient matrix formalism \cite{ADNKVrt2020}. 
Based on a simulated diverging beam reflection model, we first formulate an exact mathematical description of the reflected field profile that completely characterizes the inherent polarization inhomogeneity of the reflected field 
(a collimated final output field in 2D).
We then utilize this complete field information to explore complex optical singularity dynamics in the beam field. 
Two interrelated but distinctly characterized classes of generic optical singularities are phase and polarization singularities \cite{Gbur}. In a 2D beam cross section, a phase singularity is a point with an indeterminate phase of the beam field \cite{NyeBerry1974, Bhandari97, Soskin1997, PA2000, SV2001, DOP09, BNRev, Gbur}; whereas, a polarization singularity is a point/line with atleast one indeterminate property of the polarization ellipse (e.g., a $C$-point singularity is an isolated circular-polarization point with an indeterminate `ellipse orientation') \cite{Gbur, BerryHannay1977, Nye83b, Nye83a, Hajnal87a, Hajnal87b, NH1987, DH1994, SV2001, DOP09, DennisPS02, DennisMonstar08, NKVMonstar, NKVFiber, Vpoint}.
We impose appropriate conditions on the presently considered system to observe the phase singularity formation in a spin-polarized component field, and an associated polarization singularity formation in the total field. The degeneracy of the normal incidence gives rise to a second order phase singularity, and a corresponding `center' polarization-singular pattern \cite{Gbur}. 
We then deviate from the central singularity condition via controlled variations of the input polarization --- thus splitting the second order phase singularity to a pair of first order singularities, and the polarization singularity to a pair of $C$-point singularities with connected `lemon' polarization patterns \cite{BerryHannay1977, Gbur}. A significant singularity dynamics is thus observed via controlled variations of the system.

Subsequently, we experimentally observe the singularity dynamics by using an appropriate setup. Our simulated system and experimental setup designs, 
though are different considering the practicality of the experiment, are equivalent from the perspective of wavevector distribution in momentum space.
So, the experimentally observed phenomena show similar behaviors as those demonstrated in the simulation, thus verifying the generality and fundamental nature of the discussed singularity dynamics.


The association of spatially varying phase functions to the spin-polarized component fields imply that each spin-component field carries its own OAM. 
This is a manifestation of spin-orbit interaction (SOI) in the system \cite{AllenOAM1992, Liberman, Berry1998, Barnett2002, Onoda, Bliokh2006, Bliokh2007, HostenKwiat, Bliokh2008, Bliokh2009, BA2010, BekshaevRev2011, Qin2011, BNRev, XieSHELinIF, BliokhRev2019, BerryShukla2019, OAMBook2013}. 
The SOI originally occurs due to the inhomogeneous reflection process of the composite optical beam, 
establishing the coupling between SAM and OAM. But its signature is retained in the subsequent collimated beam due to AM conservation, and is manifested as the coupling between the spin and orbital characteristics. We use Barnett's AM flux density formalism \cite{Barnett2002} to explore these orbital characteristics by quantifying the OAM fluxes associated to the above-mentioned phase profiles of the spin-component fields. As the controlled variation of the input polarization alters the phase profiles giving rise to an interesting singularity dynamics, the associated OAM fluxes vary correspondingly, restructuring the SOI. We explore these OAM flux properties in detail, along with discussing the single-photon interpretation of a non-separable state representation of the beam field.


We thus present here a detailed exploration of the optical singularity dynamics and the associated SOI characteristics in a normal-reflected beam field. Due to the availability of the complete field information, further variations of the optical system can be performed --- potentially leading to many other interesting phenomena. The availability of exact OAM information is anticipated to be immensely useful in particle trapping and rotation, near-field analysis, nano probing and other nano-optical applications \cite{NanoOpticsBook}.


\section{The Simulated Optical System} \label{Sec_SimSys}

\subsection{A Brief Context} \label{Subsec_SimContext}

\begin{figure*}
\includegraphics[width = \linewidth]{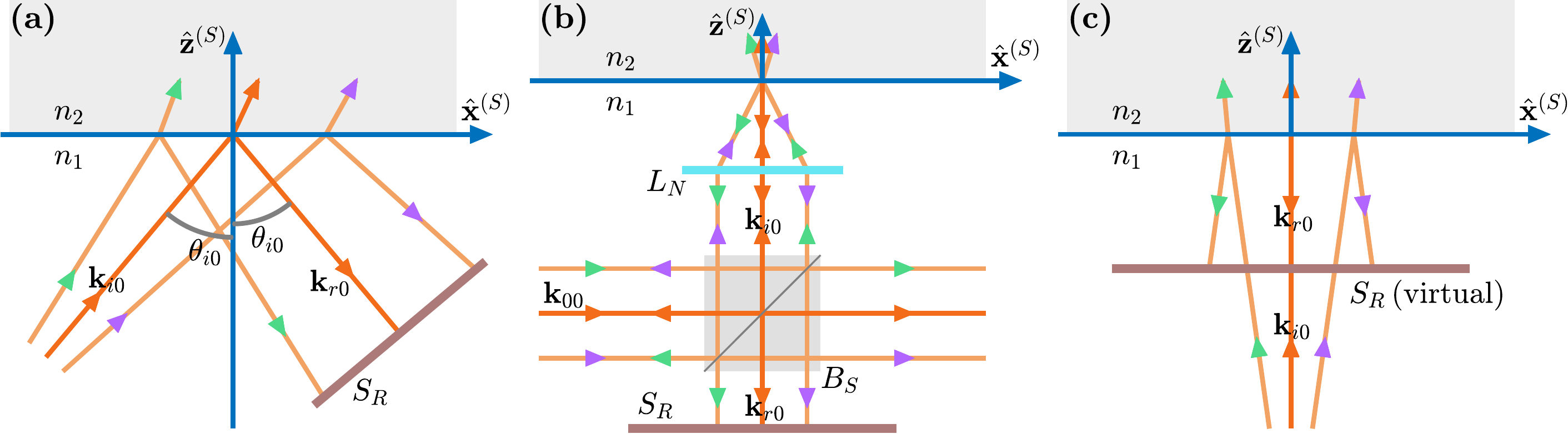}
\caption{Example schemes of optical systems for beam reflection simulations and experiments: \textbf{(a)} A diverging beam with a non-zero central angle of incidence $\theta_{i0}$; \textbf{(b)} A normally incident focused beam (standard experimental setup); \textbf{(c)} A normally incident diverging beam and a virtual screen. Here, $\mathbf{k}_{i0}$ and $\mathbf{k}_{r0}$ are respectively the incident and reflected central wavevectors; the $z^{(S)} = 0$ surface is the interface separating the dielectric media of refractive indices $n_1$ and $n_2$; and $S_R$ is the screen. In (b), $\mathbf{k}_{00}$ is the initial wavevector; $B_S$ is a beam-splitter; and $L_N$ is a converging lens that focuses the beam at $z^{(S)} = 0$.}
{\color{grey}{\rule{\linewidth}{1pt}}}
\label{Fig_SystemComp}
\end{figure*}

A reflection analysis and simulation for a general non-zero angle of incidence (of the central wavevector) can be easily performed by using a diverging beam model as shown in Fig. \ref{Fig_SystemComp}(a). 
Such a model is especially convenient for our ray-tracing analysis and simulation purposes \cite{ADNKVrt2020} as compared to a focused beam model, 
because the point-like focus formation due to geometrical ray-tracing is avoided in a diverging beam model.
The constituent wavevectors in a diverging and a converging beam model, though arranged differently in position space, occupy equivalent regions in momentum space. So, a Fresnel reflection analysis of these two composite beams are physically equivalent.

However, the system of Fig. \ref{Fig_SystemComp}(a) cannot be experimentally realized for a normal incidence case, because the screen would hinder the beam incidence. A standard normal incidence analysis usually follows the scheme of Fig. \ref{Fig_SystemComp}(b) that includes a beam-splitter to extract the reflected beam for observation \cite{NovotnyFocusImage2001, PetrovFocusedBeamRT2005, NanoOpticsBook, HerreraNanoprobing2010, FanDarkFieldMicro2012, UB_NK_NKV_OL_2022}. 
Nevertheless, there is no forbiddance to simulate normal incidence using the system of Fig. \ref{Fig_SystemComp}(a), if the screen is considered as a virtual screen [Fig. \ref{Fig_SystemComp}(c)]. 
So, for the purpose of the present paper, we present the simulated results considering the scheme of Fig. \ref{Fig_SystemComp}(c); and subsequently present the experimental results by using the standard scheme of Fig. \ref{Fig_SystemComp}(b). The equivalence of our two sets of results automatically verifies the validity of this approach.


\subsection{The System}

\begin{figure}
\includegraphics[width = 0.8\linewidth]{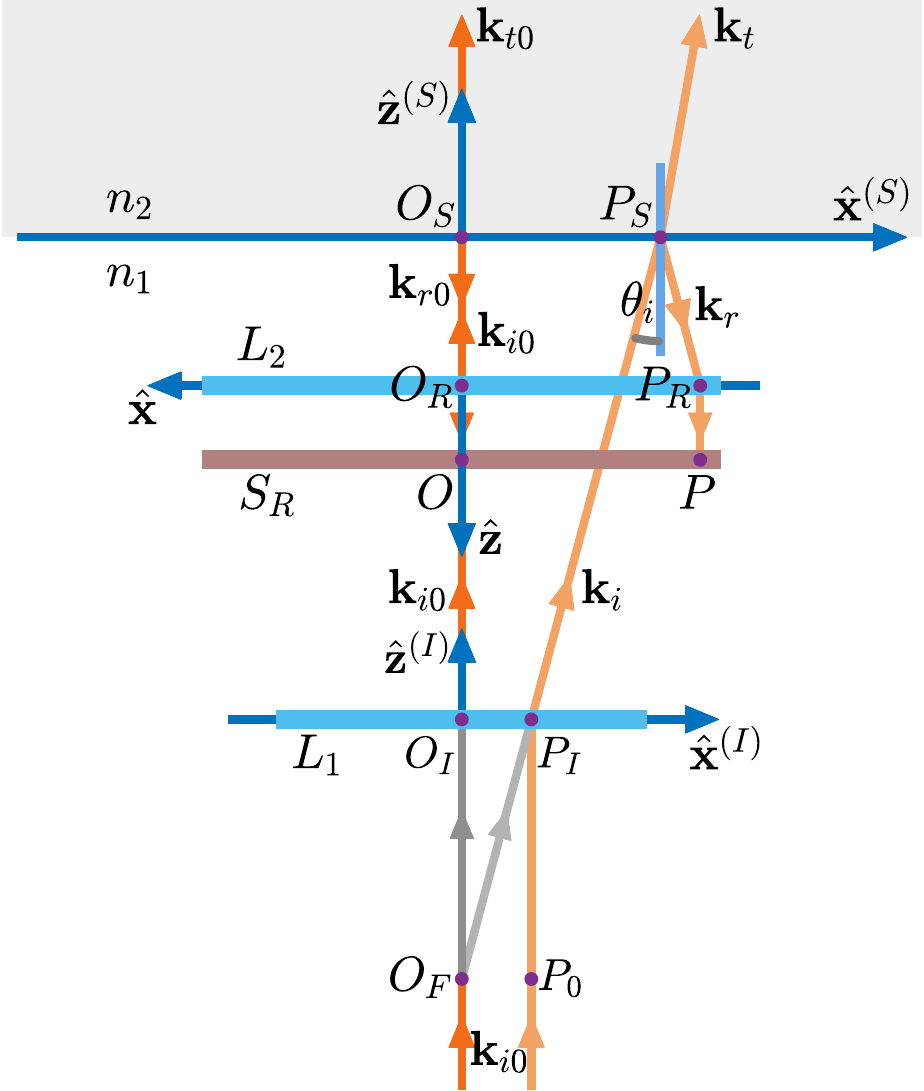}
\caption{The simulated optical system to analyze the normal incidence and reflection of a diverging optical beam (description in the text). The incident beam coordinate system $I(x^{(I)}, y^{(I)}, z^{(I)})$, the dielectric interface coordinate system $S(x^{(S)}, y^{(S)}, z^{(S)})$ and the reflected beam coordinate system $R(x, y, z)$ are related as follows: 
(1) $-\hat{\mathbf{z}}^{(S)} = $ outward normal to the interface; (2) $\hat{\mathbf{z}}^{(I)} \parallel \mathbf{k}_{i0}$, the central incident wavevector direction; (3) $\hat{\mathbf{z}} \parallel \mathbf{k}_{r0}$, the central reflected wavevector direction; (4) $\hat{\mathbf{y}}^{(I)} = \hat{\mathbf{y}}^{(S)} = \hat{\mathbf{y}}$; (5) Due to normal incidence, $\hat{\mathbf{x}}^{(I)} = \hat{\mathbf{x}}^{(S)} = -\hat{\mathbf{x}}$ and $\hat{\mathbf{z}}^{(I)} = \hat{\mathbf{z}}^{(S)} = -\hat{\mathbf{z}}$.
For the simulation we have considered a virtual collimating lens $L_2$ and a virtual observation screen $S_R$.}
{\color{grey}{\rule{\linewidth}{1pt}}}
\label{Fig_System}
\end{figure}

A complete optical system based on the scheme of Fig. \ref{Fig_SystemComp}(c) is shown in Fig. \ref{Fig_System}, where 
a collimated beam is (1) first diverged through a lens $L_1$ (with focus $O_F$, focal length $\mathcal{F}_1 = -O_F O_I$), (2) then reflected at the dielectric interface between the media of refractive indices $n_1$ and $n_2$, (3) subsequently collimated by a virtual lens $L_2$ (focal length $\mathcal{F}_2 = O_F O_I + O_I O_S + O_S O_R$), and finally (4) observed at a virtual screen $S_R$. 


Since the initial and final beams are collimated,
it is sufficient to perform our analysis based only on the electric field amplitude vector profiles, by suppressing all $\mathbf{k} \cdot \mathbf{r} - \omega t$ phase terms.
We consider an initial collimated Gaussian beam-field profile, as a function of $(x^{(I)}, y^{(I)})$ in the $I$ coordinate system [Fig. \ref{Fig_System}], as
\begin{subequations} \label{E0I_full}
\begin{eqnarray}
& \boldsymbol{\mathcal{E}}_0^{(I)} 
= \boldsymbol{\mathcal{E}}_{0x}^{(I)} + e^{i\Phi_E} \boldsymbol{\mathcal{E}}_{0y}^{(I)} = \mathcal{E}_{0x}^{(I)} \hat{\mathbf{x}}^{(I)} + e^{i\Phi_E} \mathcal{E}_{0y}^{(I)} \hat{\mathbf{y}}^{(I)}; & \label{E0I} \\
& \mathcal{E}_{0x}^{(I)} = \mathcal{E}_{00} \, G_I \cos\theta_E , \hspace{1em} \mathcal{E}_{0y}^{(I)} = \mathcal{E}_{00} \, G_I \sin\theta_E; & \label{E0x,E0y} \\
& G_I = e^{-\rho^{(I)\, 2}/w_0^2}, \hspace{1em} \rho^{(I)} = \left( x^{(I)\,2} + y^{(I)\,2} \right)^\frac{1}{2}. & \label{GI,rhoI}
\end{eqnarray}
\end{subequations}
where, $\mathcal{E}_{00} = $ central field magnitude; $w_0 = $ half beam-width; $(\theta_E, \Phi_E) = $ angle and relative-phase parameters to determine the polarization of $\boldsymbol{\mathcal{E}}_{0}^{(I)}$. 
The objective here is to obtain the final output field at the screen $S_R$. For this purpose, we first decompose 
the initial input beam into constituent rays, along each of which a family of wavefront-surface-elements are considered to propagate. We then calculate the evolution of the field at these surface-elements by tracing a complete ray-path of the form $P_0 \! \rightarrow \! P_I \! \rightarrow \! P_S \! \rightarrow \! P_R \! \rightarrow \! P$, as shown in Fig. \ref{Fig_System}. By collecting the field information for each such end point $P$, we finally obtain the complete field information at the screen $S_R$. The relevant derivation steps are shown in the Appendix.

The above procedure is capable of computationally producing exact field information even for a highly diverging beam and for any general central angle of incidence $\theta_{i0}$ \cite{ADNKVrt2020}, even though it is difficult in general to obtain an exact final analytical expression.
However, owing to the relative simplicity of the presently considered normal incidence case ($\theta_{i0} = 0^\circ$), the final field expression at the screen $S_R$ is exactly calculable --- which we have obtained as a function of $(x,y)$ in the $R$ coordinate system [Fig. \ref{Fig_System}] as
\begin{equation}
\boldsymbol{\mathcal{E}} = \boldsymbol{\mathcal{E}}^{X} + e^{i\Phi_E} \boldsymbol{\mathcal{E}}^{Y} ; \hspace{0.8em} \boldsymbol{\mathcal{E}}^{Q} = \mathcal{E}_x^{Q} \, \hat{\mathbf{x}} + \mathcal{E}_y^{Q} \, \hat{\mathbf{y}} , \; (Q = X, Y); \label{E(R)_final}
\end{equation}
where, $\boldsymbol{\mathcal{E}}^{X}$ and $\boldsymbol{\mathcal{E}}^{Y}$ are the individual outputs corresponding to the component input fields $\boldsymbol{\mathcal{E}}_{0x}^{(I)}$ and $\boldsymbol{\mathcal{E}}_{0y}^{(I)}$ respectively [Eq. (\ref{E0I})], defined by ($c_E = \cos\theta_E$, $s_E = \sin\theta_E$)
\begin{subequations}
\label{ExyXY_comps}
\begin{eqnarray}
& \mathcal{E}_x^{X} = C_1 c_E \left( 1 - C_2 u^2 \right) ; \hspace{1em} \mathcal{E}_y^{X} = - C_1 C_2 c_E u v \, ; & \\ 
& \mathcal{E}_x^{Y} = C_1 C_2 s_E u v \, ; \hspace{1em} \mathcal{E}_y^{Y} = - C_1 s_E \left( 1 - C_2 v^2 \right) ; & 
\end{eqnarray}
\end{subequations}
with the various terms defined as
\begin{subequations}
\label{terms_def}
\begin{eqnarray}
& u = x/\mathcal{F}_2, \hspace{1em} v = y/\mathcal{F}_2, \hspace{1em} \rho = \left(x^2 + y^2\right)^\frac{1}{2} \! , & \label{subeq_uvrho} \\
& \alpha = \mathcal{F}_2/|\mathcal{F}_1|, \hspace{1em} w_R = \alpha w_0, \hspace{1em} G_R = e^{-\rho^{2}/w_R^2}, & \label{subeq_alphawG} \\
& \sigma = \rho/\mathcal{F}_2, \hspace{0.5em} N = n_2^2/n_1^2 - 1, \hspace{0.5em} \beta = \left( n_2^2/n_1^2 + N\sigma^2 \right)^\frac{1}{2} \! , \hspace{1em} & \label{subeq_sigmaNbeta} \\
& C_1 = \mathcal{E}_{00} G_R(\beta - 1)/[\alpha(\beta + 1)], \hspace{1em} C_2 = 2/ \! \left(\sigma^2 + \beta\right) \! . \hspace{1.5em} & \label{subeq_C1C2}
\end{eqnarray}
\end{subequations}

While the initial input component fields $\boldsymbol{\mathcal{E}}_{0x}^{(I)}$ and $\boldsymbol{\mathcal{E}}_{0y}^{(I)}$ [Eq. (\ref{E0I})] are respectively transverse-magnetic (TM) and transverse-electric (TE), their individual outputs $\boldsymbol{\mathcal{E}}^{X}$ and $\boldsymbol{\mathcal{E}}^{Y}$ [Eq. (\ref{E(R)_final})] are neither. This occurs due to the complexity of the inhomogeneous reflection process at the dielectric interface. Even though 
the lens $L_2$ creates a final collimated output, the inhomogeneous nature of the reflection alters the field characteristics fundamentally, so that the polarization inhomogeneity is retained even in the collimated output beam.

However, the decomposition of $\boldsymbol{\mathcal{E}}$ [Eq. (\ref{E(R)_final})] in terms of complex TM ($\hat{\mathbf{x}}$) and TE ($\hat{\mathbf{y}}$) component fields is straightforward:
\begin{subequations}\label{E_ExEy_full}
\begin{eqnarray}
& \boldsymbol{\mathcal{E}} = \boldsymbol{\mathcal{E}}_x + \boldsymbol{\mathcal{E}}_y = \mathcal{E}_x \, \hat{\mathbf{x}} + \mathcal{E}_y \, \hat{\mathbf{y}} , \hspace{1em} \mathcal{E}_q = a_q + i \, b_q; & \label{E=Ex+Ey} \\
& a_q = \mathcal{E}_q^X + \mathcal{E}_q^Y \cos\Phi_E , \hspace{0.8em} b_q = \mathcal{E}_q^Y \sin\Phi_E ; & \label{aq,bq_def}
\end{eqnarray}
\end{subequations}
($q = x,y$). Subsequently, by writing $ \hat{\mathbf{x}} = (\hat{\boldsymbol{\sigma}}^+ + \hat{\boldsymbol{\sigma}}^-)/\sqrt{2} $ and $ \hat{\mathbf{y}} = (\hat{\boldsymbol{\sigma}}^+ - \hat{\boldsymbol{\sigma}}^-)/\sqrt{2}i $,
the decomposition of $\boldsymbol{\mathcal{E}}$ in terms of complex spin-component fields is obtained as
\begin{subequations}\label{sigmaPM_fields}
\begin{eqnarray}
& \boldsymbol{\mathcal{E}} = \boldsymbol{\mathcal{E}}_+ + \boldsymbol{\mathcal{E}}_- = \mathcal{E}_+ \, \hat{\boldsymbol{\sigma}}^+ + \mathcal{E}_- \, \hat{\boldsymbol{\sigma}}^- , \hspace{1em} \mathcal{E}_\pm = a_\pm + i \, b_\pm ; \hspace{1em} & \label{E=Ep+Em} \\
& a_{\pm} = (a_x \pm b_y)/\sqrt{2} \, , \hspace{0.8em} b_{\pm} = (b_x \mp a_y)/\sqrt{2} \, . \label{apm,bpm_def} &
\end{eqnarray}
\end{subequations}

Equations (\ref{E(R)_final})--(\ref{sigmaPM_fields}) contain the complete information on the presently considered inhomogeneously polarized final output field. 
No assumption or approximation on the divergence of the incident beam (output of $L_1$) is made while deriving these equations. So these exact expressions are applicable to all divergences in the paraxial as well as non-paraxial regimes. 
In the next section we establish a graphical approach to understand and extract the field information contained in the above equations.


\section{The Nature of Polarization Inhomogeneity} \label{Sec_PolInhomogeneity}


\begin{figure*}
\begin{center}
\includegraphics[width = \linewidth]{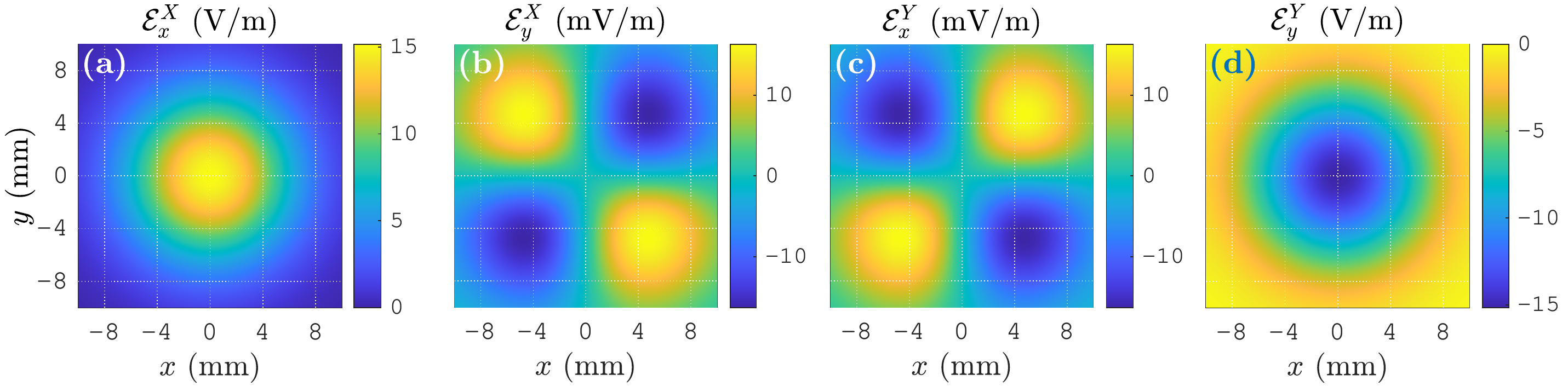}
\end{center}
\caption{The component field profiles $\mathcal{E}_x^X$, $\mathcal{E}_y^X$, $\mathcal{E}_x^Y$ and $\mathcal{E}_y^Y$ [Eqs. (\ref{ExyXY_comps})] as functions of $(x,y)$ at the output beam cross-section, for $\theta_E = 45^\circ$, arbitrary $\Phi_E$ and the considered simulation parameters.}
{\color{grey}{\rule{\linewidth}{1pt}}}
\label{Fig_ExyXY_Profiles}
\end{figure*}

For the simulations in the present paper, we consider the following parameter values: refractive indices $n_1 = 1$, $n_2 = 1.52$; laser power $\mathcal{P}_0 = 1$ mW; free-space wavelength $\lambda = 632.8$ nm; half-width of the input beam, $w_0 = 2$ mm; focal lengths of the lenses, $\mathcal{F}_1 = -3$ cm, $\mathcal{F}_2 = 10$ cm; propagation distances $O_I O_S = 5$ cm, $O_S O_R = 2$ cm \footnote{\textbf{Note:} In this example, the divergence angle is $\Theta_D = -\tan^{-1}(w_0/\mathcal{F}_1) \approx 3.814^\circ$, which lies in the paraxial regime. However, we have obtained consistent simulation results for higher divergences in the non-paraxial regime as well. One importance of obtaining simulated results in the paraxial regime is to establish the understanding that, no matter how small the divergence is, the singularity dynamics take place as long as the incident wave is not an ideal plane wave.}. 
The $\mathcal{E}_x^X$, $\mathcal{E}_y^X$, $\mathcal{E}_x^Y$ and $\mathcal{E}_y^Y$ profiles [Eqs. (\ref{ExyXY_comps})] for $\theta_E = 45^\circ$ (and arbitrary $\Phi_E$) are shown in Fig. \ref{Fig_ExyXY_Profiles}. 
As understood from Eqs. (\ref{ExyXY_comps}), $\mathcal{E}_x^X$ and $\mathcal{E}_y^Y$ are dominant field terms (approximately Gaussian form); whereas $\mathcal{E}_y^X$ and $\mathcal{E}_x^Y$ are second order terms (four-lobe form due to $uv$) representing small corrections to the field. This nature is clearly understood by observing the differences in the orders of magnitudes of the profiles of Fig. \ref{Fig_ExyXY_Profiles}. 
We can thus interpret that the input field $\boldsymbol{\mathcal{E}}_{0x}^{(I)}$ [Eq. (\ref{E0I})] creates the dominant field $\mathcal{E}_x^X \, \hat{\mathbf{x}}$ and the orthogonally polarized `correction' field $\mathcal{E}_y^X \, \hat{\mathbf{y}}$; whereas, the input field $\boldsymbol{\mathcal{E}}_{0y}^{(I)}$ [Eq. (\ref{E0I})] creates the dominant field $\mathcal{E}_y^Y \, \hat{\mathbf{y}}$ and the orthogonally polarized `correction' field $\mathcal{E}_x^Y \, \hat{\mathbf{x}}$. This form of expressing the fields is comparable to the dominant-remnant field superposition described in Ref. \cite{ADNKVGenVortex2022}.

In the complete input field $\boldsymbol{\mathcal{E}}_0^{(I)} $ [Eq. (\ref{E0I})], the component field $\boldsymbol{\mathcal{E}}_{0y}^{(I)}$ has a phase-lead of $\Phi_E$ over the component field $\boldsymbol{\mathcal{E}}_{0x}^{(I)}$ --- and hence this phase difference also appears between $\boldsymbol{\mathcal{E}}^{X}$ and $\boldsymbol{\mathcal{E}}^{Y}$ [Eq. (\ref{E(R)_final})]. So, in terms of the profiles of Fig. \ref{Fig_ExyXY_Profiles}, $\mathcal{E}_x^Y$ superposes with $\mathcal{E}_x^X$ with a phase-lead $\Phi_E$ to give a complex $\hat{\mathbf{x}}$-polarized field $\boldsymbol{\mathcal{E}}_x$; whereas $\mathcal{E}_y^Y$ superposes with $\mathcal{E}_y^X$ with a phase-lead $\Phi_E$ to give a complex $\hat{\mathbf{y}}$-polarized field $\boldsymbol{\mathcal{E}}_y$ [Eqs. (\ref{E_ExEy_full})]. The superposition of $\boldsymbol{\mathcal{E}}_x$ and $\boldsymbol{\mathcal{E}}_y$ then gives the complete field $\boldsymbol{\mathcal{E}}$ [Eq. (\ref{E=Ex+Ey})], which has a very complicated form due to the above-mentioned effects.

The above discussion completely reveals the true nature of the polarization inhomogeneity observed in the final output beam field. Though we discuss this inherent nature in the presently considered normal incidence case based on Eqs. (\ref{ExyXY_comps}) and the profiles of Fig. \ref{Fig_ExyXY_Profiles}, we have verified based on the formalism of Ref. \cite{ADNKVrt2020} that the generic forms of Eqs. (\ref{E(R)_final}), (\ref{E_ExEy_full}) and (\ref{sigmaPM_fields}) seamlessly apply to all cases of general central angles of incidence $\theta_{i0}$. 
The generation of optical singularities in Refs. \cite{ADNKVBrew2021} and \cite{ADNKVGenVortex2022}, for example, relies on this generic form of field decomposition. 
This inherent inhomogeneous polarization of the field is attained originally due to the composite beam reflection at the dielectric interface; and hence contains all the information necessary for understanding the beam-shift and spin-shift phenomena. For the purpose of the present paper, we explore specific details of the optical singularities and the relevant OAM characteristics by studying this polarization inhomogeneity of the beam field.



\section{Optical Singularity Dynamics} \label{Sec_SingularityDynamics}

\subsection{Central Phase Singularity} \label{Subsec_CentralSingularity}

\begin{figure*}
\begin{center}
\includegraphics[width = \linewidth]{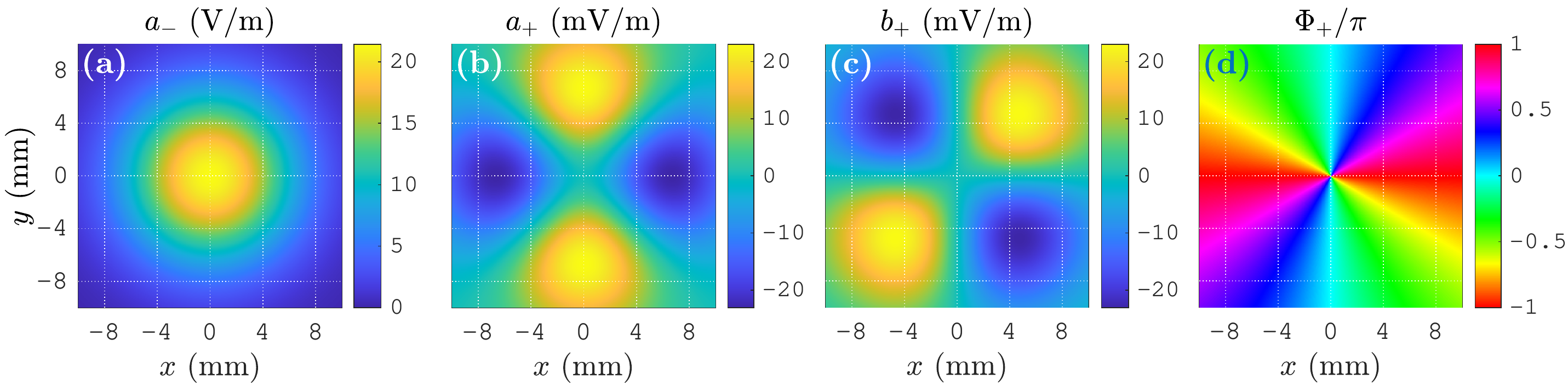}
\end{center}
\caption{Profiles of the functions $a_\pm$, $b_+$ [Eqs. (\ref{abpm_specialCase_full})] and $\Phi_+$ [Eq. (\ref{PhiPlus_def})] as functions of $(x,y)$ at the output beam cross-section, for $\theta_E = 45^\circ$ and $\Phi_E = +\pi/2$.}
{\color{grey}{\rule{\linewidth}{1pt}}}
\label{Fig_abpmPhp_Profiles}
\end{figure*}

In an inhomogeneously polarized beam field, if an isolated point $P$ exists where $\boldsymbol{\mathcal{E}}_x$ and $\boldsymbol{\mathcal{E}}_y$ have equal magnitudes, then setting a phase difference $\pi/2$ between them creates an isolated circular polarization, i.e. a $C$-point singularity, at $P$ \cite{ADNKVGenVortex2022} (the only exception occurs if $|\boldsymbol{\mathcal{E}}_x|_P = |\boldsymbol{\mathcal{E}}_y|_P = 0$, creating a higher order singularity \cite{ADNKVBrew2021}). If the $C$-point is $\hat{\bs{\mathcal{\sigma}}}^+$ ($\hat{\bs{\mathcal{\sigma}}}^-$) polarized, then the component field $\boldsymbol{\mathcal{E}}_+$ ($\boldsymbol{\mathcal{E}}_-$) dominates in the neighborhood with a uniform/near-uniform phase profile; whereas, the less intense (`remnant' \cite{ADNKVGenVortex2022}) field $\boldsymbol{\mathcal{E}}_-$ ($\boldsymbol{\mathcal{E}}_+$) manifests a phase singularity at that point.
In particular, for the presently considered normal incidence case, it is understood from Eqs. (\ref{E0I_full})--(\ref{sigmaPM_fields}) that the choice of $\theta_E = \pm45^\circ$ and $\Phi_E = \pm\pi/2$ can generate such a singularity precisely at the beam-center. 
For the present paper, it is sufficient to demonstrate the case of $\theta_E = 45^\circ$ and $\Phi_E = +\pi/2$. The cases involving $\theta_E = -45^\circ$ and $\Phi_E = -\pi/2$ can be worked out in the same way, and need not be discussed explicitly.


For the considered case of $(\theta_E, \Phi_E) = (45^\circ, +\pi/2)$, the initial input field $\boldsymbol{\mathcal{E}}_0^{(I)} $ [Eq. (\ref{E0I})] is purely $\hat{\boldsymbol{\sigma}}^+$ spin-polarized; and hence, the $\hat{\boldsymbol{\sigma}}^-$ spin-polarized component field $\boldsymbol{\mathcal{E}}_-$ [Eq. (\ref{E=Ep+Em})] is the dominant output (signifying a spin-flip due to reflection). This implies that the intended phase singularity is to be observed in the phase of the less dominant $\hat{\boldsymbol{\sigma}}^+$ spin-polarized output field $\boldsymbol{\mathcal{E}}_+$.
In the light of the discussions of Section \ref{Sec_PolInhomogeneity}, we now establish a graphical understanding of the $\boldsymbol{\mathcal{E}}_\pm$ fields [Eqs. (\ref{sigmaPM_fields})] in order to understand the phase singularity formation.

By using $(\theta_E, \Phi_E) = (45^\circ, +\pi/2)$ in Eqs. (\ref{ExyXY_comps}), (\ref{aq,bq_def}) and (\ref{apm,bpm_def}), we obtain
\begin{subequations}\label{abpm_specialCase_full}
\begin{eqnarray}
& a_- = C_1 - (C_1 C_1/2) \, \sigma^2 , \hspace{1em} b_- = 0, \hspace{0.5em} & \\
& a_+ = (C_1 C_2 / 2) \left(v^2 - u^2\right), \hspace{1em} b_+ = (C_1 C_2 / 2) \, 2 u v. \hspace{0.5em} & \label{abp_specialCase}
\end{eqnarray}
\end{subequations}
These $a_-$, $a_+$ and $b_+$ profiles, as functions of $(x,y)$, are shown in Figs. \ref{Fig_abpmPhp_Profiles}(a), \ref{Fig_abpmPhp_Profiles}(b) and \ref{Fig_abpmPhp_Profiles}(c) respectively. 
The phase of $\boldsymbol{\mathcal{E}}_+$ is given by
\begin{equation}
\Phi_+ = \tan^{-1} (b_+/a_+); \label{PhiPlus_def}
\end{equation}
which, for the case of Eq. (\ref{abp_specialCase}), reduces to $\Phi_+ = \pi - 2\phi$, where $\phi$ is the azimuthal coordinate in the concerned $xy$ plane of the screen. Here, the $\pi$ term is considered in order to follow the phase convention of Ref. \cite{ADNKVrt2020}, along with limiting the phase range as $(-\pi,\pi]$. 

The variation of $\Phi_+$ satisfying $d\Phi_+/d\phi = -2$ implies that the field function $\mathcal{E}_+$ [Eq. (\ref{E=Ep+Em})] is a pure LG mode of index $l = -2$. The vortex of $\Phi_+$ can thus be classified as a canonical vortex \cite{Berry1998, OAMBook2013}.
The $\Phi_+$ profile as a function of $(x,y)$ is shown in Fig. \ref{Fig_abpmPhp_Profiles}(d), which clearly shows the $l = -2$ phase vortex at the beam-center. 
This central phase singularity is attributed to $a_+ = b_+ = 0$ at the beam-center [Figs. \ref{Fig_abpmPhp_Profiles}(b), \ref{Fig_abpmPhp_Profiles}(c)], which makes the phase indeterminate. In this way, by considering $(\theta_E, \Phi_E) = (45^\circ, +\pi/2)$, the intended phase singularity is obtained at the normal-reflected beam-center.

While the initial input field contains only spin angular momentum (SAM), the generation of the phase vortex implies that the final output field contains some amount of OAM as well. The complete analysis given by Eqs. (\ref{E(R)_final})--(\ref{PhiPlus_def}) clearly reveals that this OAM is obtained due to the inhomogeneous nature of the reflection process --- which is a definite signature of SOI, manifesting itself as a partial conversion from SAM to OAM.

\subsection{Off-Central Phase Singularities} \label{Subsec_OffCentralSingularity}

As discussed in Ref. \cite{ADNKVGenVortex2022}, shifting $\theta_E$ around the central value (here $45^\circ$) shifts the singularity to off-central positions. We explore this dynamics for the presently considered normal incidence case by considering shifted $\theta_E$ values $\theta_E = 45^\circ + \Delta\theta_E$, where $|\sin \Delta\theta_E| \ll 1$.
To obtain a singularity of $\Phi_+$ (Eq. (\ref{PhiPlus_def}), for general $a_+$ and $b_+$) at a point, we require the $\Phi_+$ value to be indeterminate at that point due to the condition $b_+ = a_+ = 0$. This condition, by virtue of Eqs. (\ref{ExyXY_comps}), (\ref{terms_def}), (\ref{aq,bq_def}) and (\ref{apm,bpm_def}), for $\Phi_E = +\pi/2$ and for a general non-zero $\Delta\theta_E$, translates to
\begin{equation}
uv = 0 \, ; \hspace{1em} \tan \Delta\theta_E = (v^2 - u^2)/\beta.
\end{equation}
Using these equations, we obtain the solutions for the singularity points, as functions of $\Delta\theta_E$, as
\begin{subequations}\label{S_coordinates}
\begin{eqnarray}
& (x_{1\pm}, y_1) = (\pm d_0, 0), \hspace{0.5em} \mbox{for} \; \Delta\theta_E \leq 0; \hspace{0.5em} & \label{(pml0,0)} \\
& (x_2, y_{2\pm}) = (0, \pm d_0), \hspace{0.5em} \mbox{for} \; \Delta\theta_E \geq 0; \hspace{0.5em} & \label{(0,pml0)} \\
& d_0 = \dfrac{\mathcal{F}_2 \tan\Delta\theta_E}{\sqrt{2}} \left[ N + \left(N^2 + \dfrac{4n_2^2}{n_1^2 \tan^2\Delta\theta_E}\right)^\frac{1}{2} \right]^\frac{1}{2} \! \! . \hspace{1.5em} & \label{l0_def}
\end{eqnarray}
\end{subequations}
We observe the dynamics of the off-central singularities based on Eqs. (\ref{S_coordinates}) by varying only $\Delta\theta_E$, but keeping $\Phi_E$ constant. So we suppress the explicit mention of $\Phi_E = +\pi/2$ here onwards, wherever appropriate.


\begin{figure*}
\begin{center}
\includegraphics[width = 0.88\linewidth]{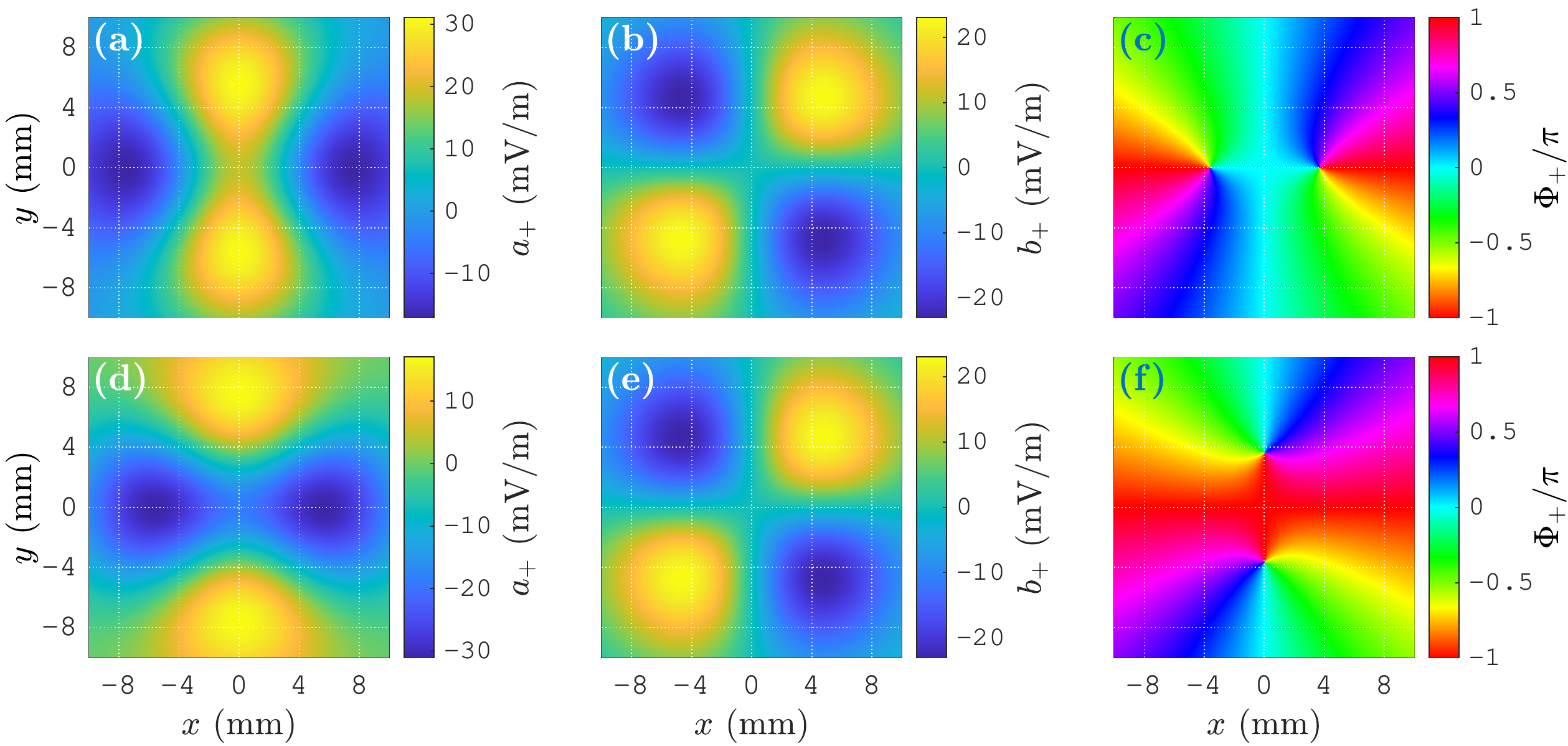}
\end{center}
\caption{Profiles of the functions $a_+$, $b_+$ [Eq. (\ref{apm,bpm_def})] and $\Phi_+$ [Eq. (\ref{PhiPlus_def})] as functions of $(x,y)$ at the output beam cross-section, for $\Phi_E = +\pi/2$ and \textbf{(a), (b), (c)} $\Delta\theta_E = -0.05^\circ$; \textbf{(d), (e), (f)} $\Delta\theta_E = +0.05^\circ$.}
{\color{grey}{\rule{\linewidth}{1pt}}}
\label{Fig_0PM_abpPhp_Profiles}
\end{figure*}

The simulated $a_+$, $b_+$ and $\Phi_+$ profiles for $\Delta\theta_E = -0.05^\circ$ (a $\Delta\theta_E < 0^\circ$ case) are shown in Figs. \ref{Fig_0PM_abpPhp_Profiles}(a), \ref{Fig_0PM_abpPhp_Profiles}(b) and \ref{Fig_0PM_abpPhp_Profiles}(c) respectively. In this case the condition $a_+ = b_+ = 0$ is satisfied at two symmetrically opposite points $(\pm d_0, 0)$ [Eq. (\ref{(pml0,0)})] on the $x$ axis --- forming the phase singularities at these two points.
Also, the simulated $a_+$, $b_+$ and $\Phi_+$ profiles for $\Delta\theta_E = +0.05^\circ$ (a $\Delta\theta_E > 0^\circ$ case) are shown in Figs. \ref{Fig_0PM_abpPhp_Profiles}(d), \ref{Fig_0PM_abpPhp_Profiles}(e) and \ref{Fig_0PM_abpPhp_Profiles}(f) respectively; for which the condition $a_+ = b_+ = 0$ is satisfied at two symmetrically opposite points $(0, \pm d_0)$ [Eq. (\ref{(0,pml0)})] on the $y$ axis --- generating the phase singularities at these two points.

Unlike the $\Delta\theta_E = 0^\circ$ case, however, the paired off-central phase vortices are not canonical vortices, because the variation of $\Phi_+$ around the vortices does not have a simple proportionality relation with the azimuthal rotation around the vortices. Such a vortex cannot be assigned an LG mode index $l$; but instead, can be described by assigning a topological charge \cite{NyeBerry1974, BH1977, OAMBook2013, Gbur}
\begin{equation}
t = \dfrac{1}{2\pi} \oint_\mathcal{C} \nabla\Phi(\mathbf{r}) \cdot d\mathbf{r},
\end{equation}
where, $\mathcal{C}$ denotes a closed contour around the vortex. The $\Phi_+$ profiles of Figs. \ref{Fig_0PM_abpPhp_Profiles}(c) and \ref{Fig_0PM_abpPhp_Profiles}(f) show that, for a counter-clockwise rotation around each of the vortices, $\Phi_+$ changes by $-2\pi$. This implies that the topological charge of each of the off-central vortices is $t = -1$. On the contrary, the topological charge of the pure LG mode $\Phi_+$ profile of Fig. \ref{Fig_abpmPhp_Profiles}(d) is identical to its mode index: $t = l = -2$. This leads us to a significant vortex dynamics, interpreted consistently in the following two ways:


\begin{enumerate}
\item In terms of Eq. (\ref{(pml0,0)}): As $\Delta\theta_E$ approaches zero from a negative value, the two $t = -1$ phase vortices situated at symmetrically opposite positions on the $x$ axis [Fig. \ref{Fig_0PM_abpPhp_Profiles}(c)] approach each other and merge at the origin to create a $t = l = -2$ phase vortex [Fig. \ref{Fig_abpmPhp_Profiles}(d)].

\item In terms of Eq. (\ref{(0,pml0)}): As $\Delta\theta_E$ approaches zero from a positive value, the two $t = -1$ phase vortices situated at symmetrically opposite positions on the $y$ axis [Fig. \ref{Fig_0PM_abpPhp_Profiles}(f)] approach each other and merge at the origin to create a $t = l = -2$ phase vortex [Fig. \ref{Fig_abpmPhp_Profiles}(d)].
\end{enumerate}
This observation thus reveals a remarkable dynamics of the phase vortices, which has not been anticipated in the earlier work of Ref. \cite{ADNKVGenVortex2022}. The SOI manifesting itself in the form of a merger of two lower order phase vortices to create a higher order phase vortex is a significant result obtained in the present work.



\subsection{Polarization Singularities} \label{Subsec_PolSing}

\begin{figure*}
\begin{center}
\includegraphics[width = 0.88\linewidth]{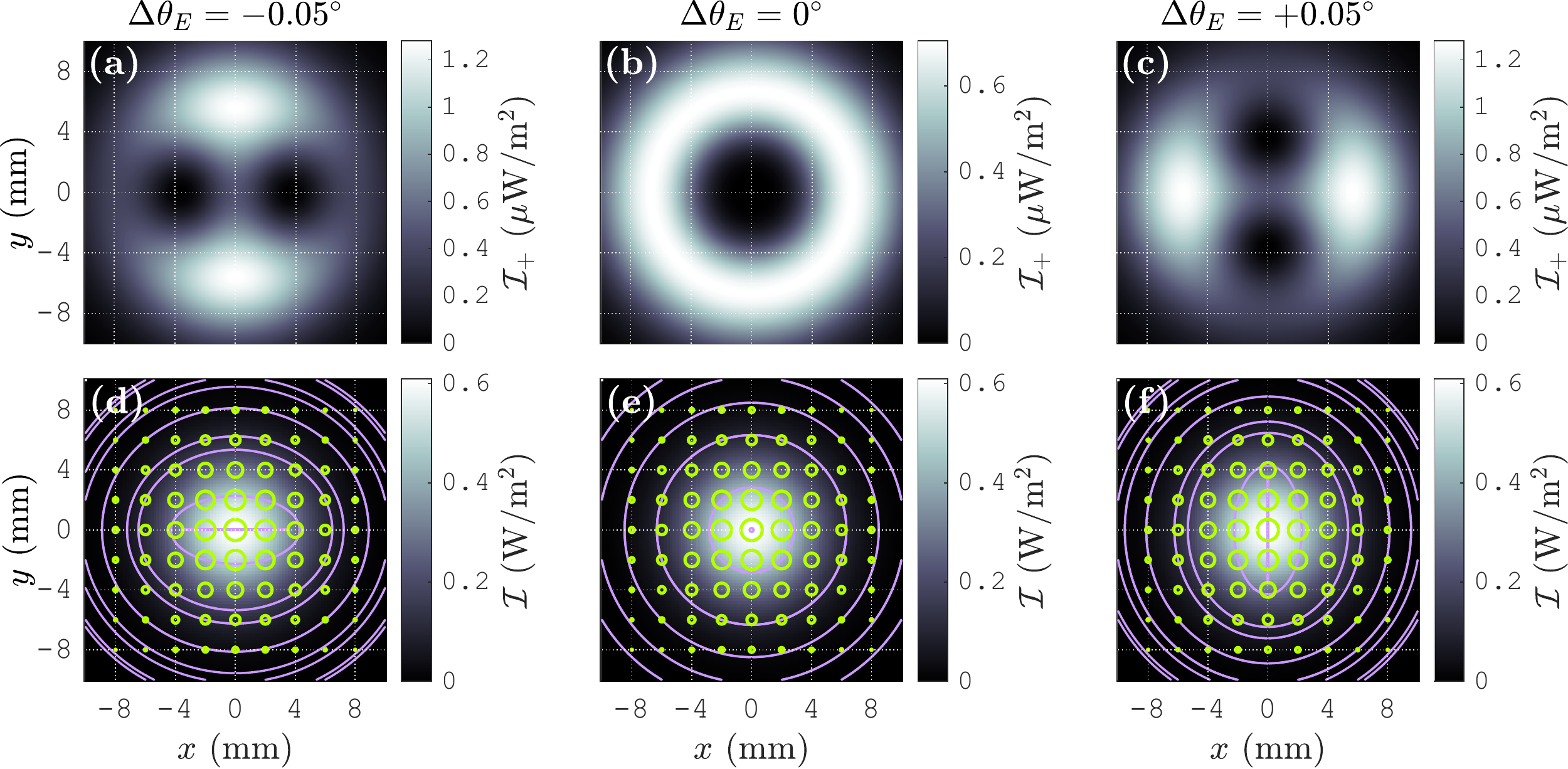}
\end{center}
\caption{\textbf{(a), (b), (c)} Simulated intensity profiles $\mathcal{I}_+ = (n_1/2\mu_0 c)|\boldsymbol{\mathcal{E}}_+|^2$ for $\Delta\theta_E = -0.05^\circ, 0^\circ, +0.05^\circ$ and $\Phi_E = +\pi/2$; \textbf{(d), (e), (f)} Corresponding total field profiles $\boldsymbol{\mathcal{E}}$ [Eqs. (\ref{E_ExEy_full}), (\ref{sigmaPM_fields})].
The ellipticities of the polarization ellipses are not visually understandable, since the $\hat{\boldsymbol{\sigma}}^-$-polarized component field $\boldsymbol{\mathcal{E}}_-$ significantly dominates over $\boldsymbol{\mathcal{E}}_+$. However, the arrangement of the polarization ellipses are understood by generating the streamlines.
}
{\color{grey}{\rule{\linewidth}{1pt}}} \vspace{-0.4em}
\label{Fig_0PM_IE_Profiles}
\end{figure*}

Due to the condition $a_+ = b_+ = 0$, the intensity of the $\boldsymbol{\mathcal{E}}_+$ field goes to zero at the singularity coordinates. From a physical perspective, a non-zero field cannot exist with an indeterminate phase. This nature is clearly observed in the intensity profiles $\mathcal{I}_+ = (n_1/2\mu_0 c)|\boldsymbol{\mathcal{E}}_+|^2$ of the field $\boldsymbol{\mathcal{E}}_+$ shown in Figs. \ref{Fig_0PM_IE_Profiles}(a), \ref{Fig_0PM_IE_Profiles}(b) and \ref{Fig_0PM_IE_Profiles}(c) for $\Delta\theta_E = -0.05^\circ, 0^\circ, +0.05^\circ$ respectively.

Since $\boldsymbol{\mathcal{E}}_+$ is exactly zero at the phase-singularity points, the total field $\boldsymbol{\mathcal{E}}$ contains pure $\hat{\boldsymbol{\sigma}}^-$ polarizations at these points due to the contribution of the $\boldsymbol{\mathcal{E}}_-$ field only [Eq. (\ref{E=Ep+Em})]. Clearly, both the fields $\boldsymbol{\mathcal{E}}_\pm$ are non-zero at the nearby points, and hence the polarizations are elliptical in the vicinity. 
Thus, isolated $\hat{\boldsymbol{\sigma}}^-$ polarizations are obtained in the total field $\boldsymbol{\mathcal{E}}$ at the phase-singularity points of $\boldsymbol{\mathcal{E}}_+$ [Eq. (\ref{S_coordinates})] --- giving rise to $C$-point polarization singularities \cite{Gbur, BerryHannay1977, Nye83b, Nye83a, Hajnal87a, Hajnal87b, NH1987, DH1994, SV2001, DOP09, DennisPS02, DennisMonstar08, NKVMonstar, NKVFiber} at these points.

The simulated total field profiles for $\Delta\theta_E = -0.05^\circ, 0^\circ, +0.05^\circ$ are shown in Figs. \ref{Fig_0PM_IE_Profiles}(d), \ref{Fig_0PM_IE_Profiles}(e) and \ref{Fig_0PM_IE_Profiles}(f) respectively; where the formation of the $C$-point singularities are observed. The streamlines show lemon polarization patterns \cite{BerryHannay1977, Gbur} on the $x$ axis and the $y$ axis in the profiles of Figs. \ref{Fig_0PM_IE_Profiles}(d) and \ref{Fig_0PM_IE_Profiles}(f) respectively, identifying the $C$-point polarization singularities that correspond to the $t = -1$ phase singularities of Figs. \ref{Fig_0PM_abpPhp_Profiles}(c) and \ref{Fig_0PM_abpPhp_Profiles}(f). A higher order `center' polarization-singular pattern \cite{Gbur} is observed in the profile of Fig. \ref{Fig_0PM_IE_Profiles}(e), that corresponds to the $t = l = -2$ phase singularity of Fig. \ref{Fig_abpmPhp_Profiles}(d).

The two lemon patterns of Fig. \ref{Fig_0PM_IE_Profiles}(d) can be interpreted as a pair of interconnected lemon patterns on a common horizontal separatrix \cite{Gbur}; and the same interconnectedness can be interpreted for the two lemon patterns of Fig. \ref{Fig_0PM_IE_Profiles}(f) considering a common vertical separatrix. With these interpretations, the formation of the center singularity pattern in the profile of Fig. \ref{Fig_0PM_IE_Profiles}(e) can be consistently understood in the following two ways:
\begin{enumerate}
\item As $\Delta\theta_E$ approaches zero from a negative value, the two horizontally separated lemon pattens approach each other along the common horizontal separatrix and merge at the origin to create a higher order center singularity pattern.
\item As $\Delta\theta_E$ approaches zero from a positive value, the two vertically separated lemon pattens approach each other along the common vertical separatrix and merge at the origin to create a higher order center singularity pattern.
\end{enumerate}
A significant polarization singularity dynamics is thus observed in the reflected inhomogeneously polarized beam field; which is intrinsically associated to the dynamics of the $\Phi_+$ phase singularities.

A phase singularity is not necessarily associated to a polarization singularity in all general cases \cite{ADNKVGenVortex2022}.
For instance, if the presently considered $\hat{\boldsymbol{\sigma}}^-$ spin-polarized field $\boldsymbol{\mathcal{E}}_-$ is replaced with a general elliptically polarized field, the total field at the phase-singularity points of $\boldsymbol{\mathcal{E}}_+$ would be elliptical --- and hence, no $C$-point singularity would be created in the total field $\boldsymbol{\mathcal{E}}$. It is thus a special nature of the presently considered polarization inhomogeneity that a polarization singularity is obtained in the total field where one of the component fields is phase-singular. 
However, the most important result of this analysis is the revelation of the interrelation between a phase-singularity merger phenomenon and a polarization-singularity merger phenomenon. As the two lower order phase vortices of Fig. \ref{Fig_0PM_abpPhp_Profiles}(c) (or Fig. \ref{Fig_0PM_abpPhp_Profiles}(f)) approach each other and merge to form the higher order phase vortex of Fig. \ref{Fig_abpmPhp_Profiles}(d), the two simple lemon patterns of Fig. \ref{Fig_0PM_IE_Profiles}(d) (or Fig. \ref{Fig_0PM_IE_Profiles}(f)) also approach each other and merge to form the higher order center singularity pattern of Fig. \ref{Fig_0PM_IE_Profiles}(e). This is a remarkable and previously unanticipated SOI phenomenon which is much more significant and fundamentally interesting than the commonly observed partial conversion from SAM to OAM.


\subsection{Singularity Trajectories}


\begin{figure}
\begin{center}
\includegraphics[width = 0.9\linewidth]{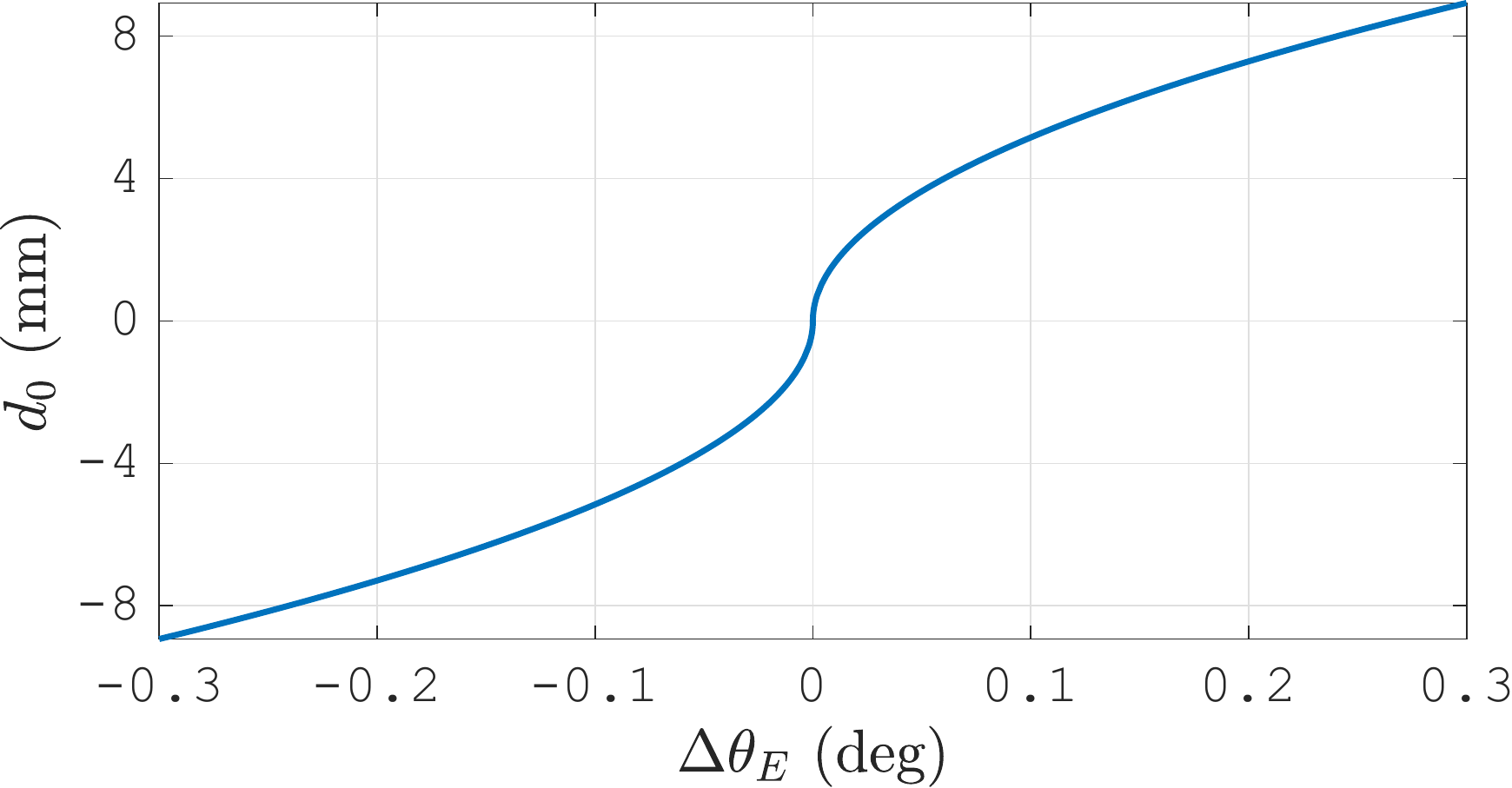}
\end{center}
\caption{Variation of $d_0$ [Eq. (\ref{l0_def})] as a function of $\Delta\theta_E$ in an example range $[-0.3^\circ,+0.3^\circ]$, for the considered simulation parameters.}
{\color{grey}{\rule{\linewidth}{1pt}}}
\label{Fig_l0_Plot}
\end{figure}

\begin{figure}
\begin{center}
\includegraphics[width = 0.8\linewidth]{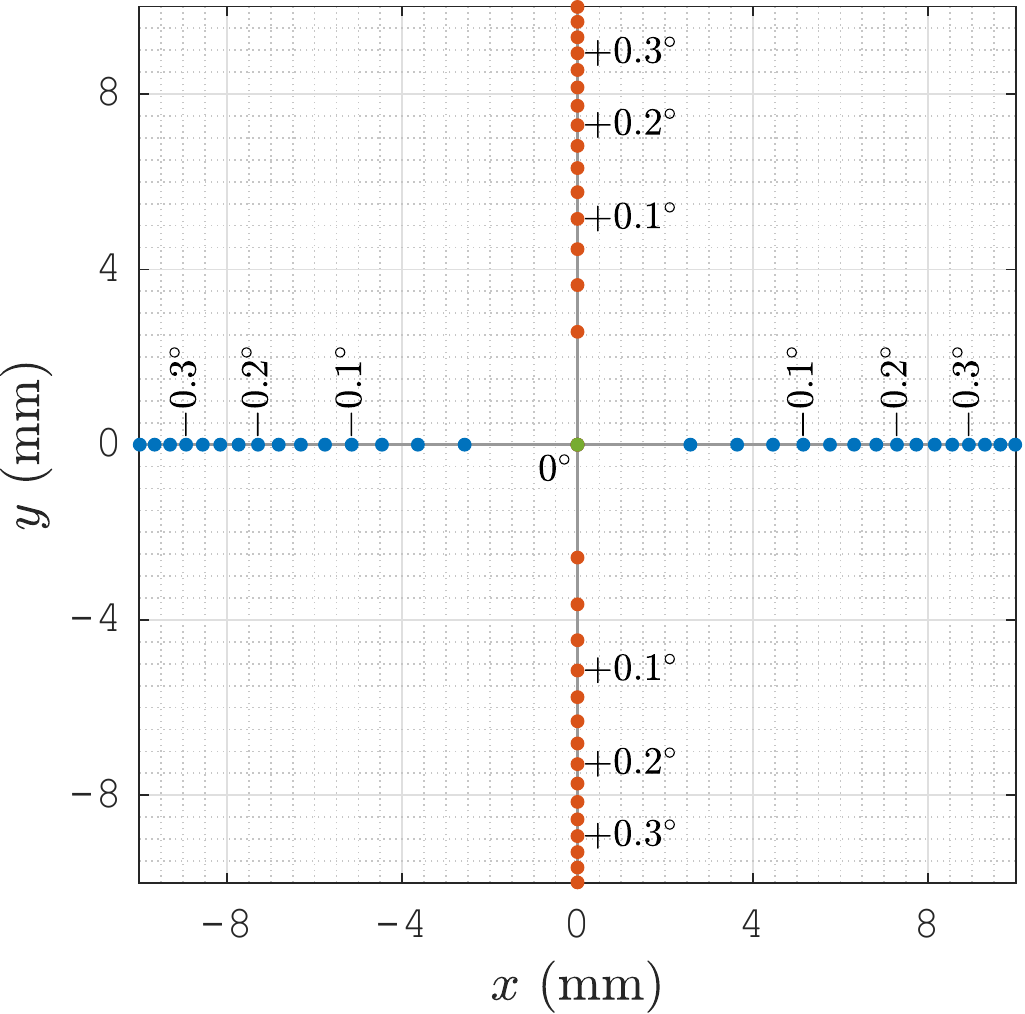}
\end{center}
\caption{Trajectories of the singularity points --- the $x$ and $y$ axes. The point-pairs of the form $(\pm d_0, 0)$ and $(0,\pm d_0)$ [Eqs. (\ref{S_coordinates})], for all $d_0$ [Eq. (\ref{l0_def}), Fig. \ref{Fig_l0_Plot}], represent the positions of the singularity-pairs for $\Delta\theta_E \leq 0$ and $\Delta\theta_E \geq 0$ respectively. Representative points are shown on these trajectories, with $\Delta\theta_E$ values marked. Here, any two adjacent points are separated by a $\Delta\theta_E$ interval of $0.025^\circ$.}
{\color{grey}{\rule{\linewidth}{1pt}}} \vspace{-0.5em}
\label{Fig_SingTrajectory}
\end{figure}

The above formulation describes well defined trajectories of the singularity points at the beam cross-section, with respect to the variation of $\Delta\theta_E$. We first observe the variation $d_0$ as a function of $\Delta\theta_E$ [Eq. (\ref{l0_def})], as shown in Fig. \ref{Fig_l0_Plot}, for the presently considered simulation parameters. Then, the coordinates $(\pm d_0, 0)$ for $\Delta\theta_E \leq 0$ and $(0,\pm d_0)$ for $\Delta\theta_E \geq 0$ [Eqs. (\ref{S_coordinates})], with $d_0$ varying according to the plot of Fig. \ref{Fig_l0_Plot}, give the trajectories of the singularity points --- as shown in Fig. \ref{Fig_SingTrajectory}. 
These trajectories are comparable to, but fundamentally different from, the displacement trajectories of non-canonical vortices which move at the beam cross-section due to beam propagation \cite{Nye83a, OAMBook2013, Gbur}.


\section{Experimental Demonstrations} \label{Sec_Exp}

\subsection{The Experimental Setup}

\begin{figure}
\includegraphics[width = \linewidth]{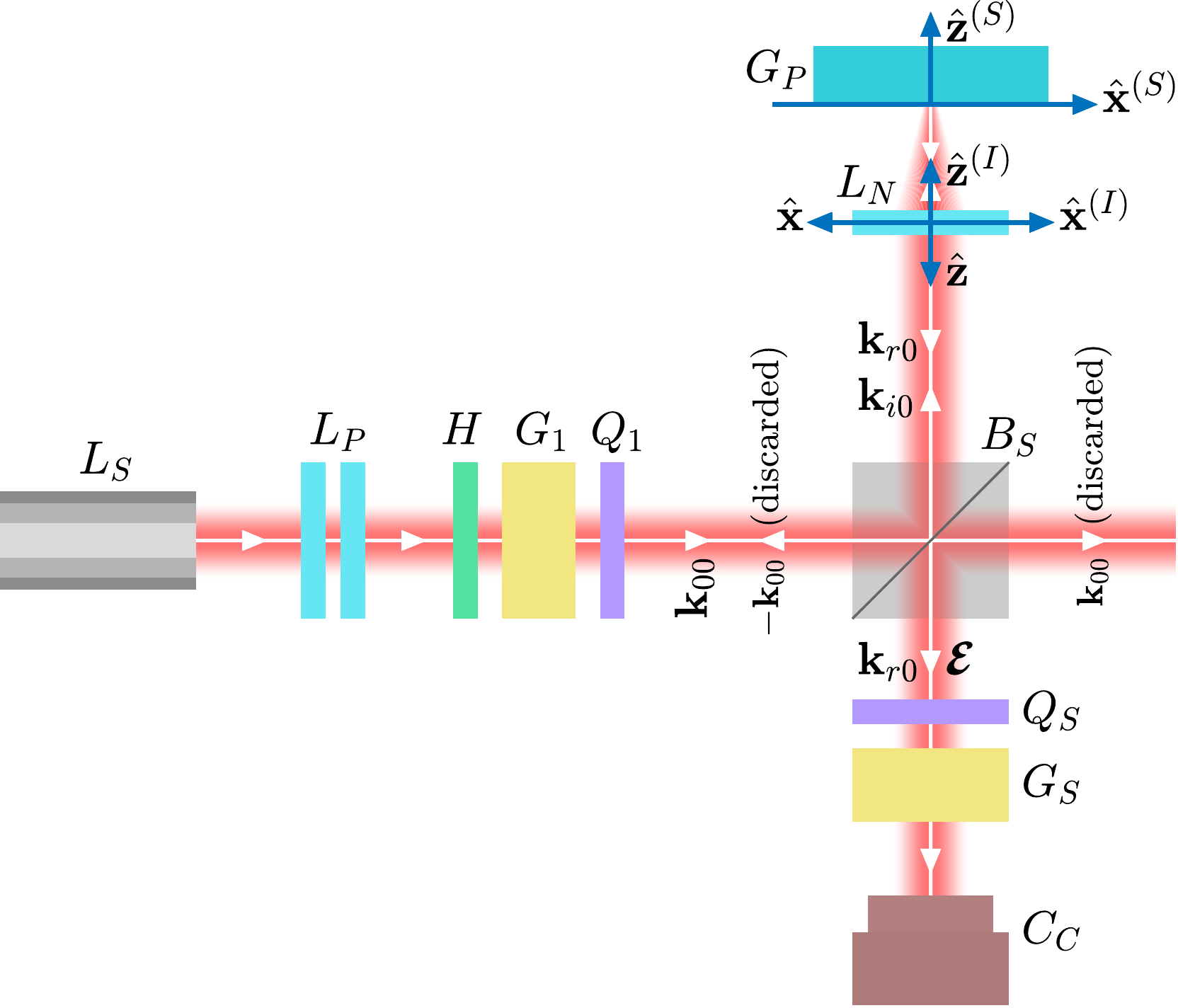}
\caption{The experimental setup, comprising the following components. $L_S$: He-Ne laser; $L_P$: Collimating lens pair; $H$: Half wave plate (HWP); $G_1, G_S$: Glan-Thompson polarizers (GTP); $Q_1, Q_S$: Quarter wave plates (QWP); $B_S$: Beam-splitter;
$L_N$: High numerical aperture (NA) lens; 
$G_P$: Glass plate (the $z^{(S)} = 0$ surface is used as the isotropic dielectric interface for reflection); $C_C$: CCD camera (as screen $S_R$). The lens $L_N$ serves the purposes of both $L_1$ and $L_2$ of Fig. \ref{Fig_System}; and hence, the coordinate systems $I(x^{(I)},y^{(I)},z^{(I)})$ and $R(x,y,z)$ are both defined with respect to the position of $L_N$.}
{\color{grey}{\rule{\linewidth}{1pt}}}
\label{Fig_ExpSetup}
\end{figure}


As discussed in Section \ref{Subsec_SimContext}, a standard setup following the scheme of Fig. \ref{Fig_SystemComp}(b) is required for a normal-reflection experiment.
Our detailed experimental setup based on this scheme is shown in Fig. \ref{Fig_ExpSetup}. We first collimate a He-Ne laser beam ($\lambda = 632.8$ nm) by using a lens pair ($L_P$), and then assign intended polarizations to it by using a combination of a half wave plate (HWP), a Glan-Thompson polarizer (GTP), and a quarter wave plate (QWP) ($H$, $G_1$ and $Q_1$). We then use a beam-splitter ($B_S$) to partially reflect the beam towards a high NA lens ($L_N$), which sharply focuses the beam at an air-glass interface. 
A microscope objective lens is used here as the high NA lens (Specifications: focal length $1.60$ mm, oil immersion NA $1.25$, magnification 100X, manufacturer Edmund Optics). 
The reflected beam hence obtained is collimated by the same lens $L_N$, and is propagated towards a CCD camera ($C_C$) after partial transmission through the beam-splitter ($B_S$). 


To observe the singularity dynamics of the field function $\mathcal{E}_+$ [Eq. (\ref{E=Ep+Em})], we extract it from the total field $\bs{\mathcal{E}}$ by using a QWP-GTP combination ($Q_S$ and $G_S$) before the camera. Orienting the fast axis of $Q_S$ along $\hat{\mathbf{y}}$ performs the transformation $\mathcal{E}_\pm \hat{\bs{\sigma}}^\pm \rightarrow \mathcal{E}_\pm \hat{\mathbf{d}}^\pm$, where $\hat{\mathbf{d}}^\pm = (\hat{\mathbf{x}} \pm \hat{\mathbf{y}})/\sqrt{2}$. Then, by orienting the transmission axis of $G_S$ along $\hat{\mathbf{d}}^+$, the field $\mathcal{E}_+ \hat{\mathbf{d}}^+$ is isolated and observed.

Considering the design difference between the simulated system [Fig. \ref{Fig_System}] and the experimental setup [Fig. \ref{Fig_ExpSetup}], we allow the length scales of the simulated and experimental profiles to be different.
In particular, the half-width of the final collimated beam ($w_R$) in the experimental setup remains the same as that of the initial collimated beam ($w_0$), which we set here as approximately $0.3$ mm. 
Additionally, the high NA lens makes the incident beam non-paraxial. This makes the field $\bs{\mathcal{E}}_+$ intense enough, thus easing the detection process of the singularities using a CCD camera. 
This also allows us to use a large enough $\Delta\theta_E$ range, thus making the measurements convenient.

It is to be noticed that, in the diverging beam model [Fig. \ref{Fig_SystemComp}(c)], a ray incident from the $x^{(S)} > 0$ (or $x^{(S)} < 0$, $y^{(S)} > 0$, $y^{(S)} < 0$) region reflects and propagates back to the $x^{(S)} > 0$ (or $x^{(S)} < 0$, $y^{(S)} > 0$, $y^{(S)} < 0$) region. On the other hand, in the focused beam model [Fig. \ref{Fig_SystemComp}(b)], a ray incident from the $x^{(S)} > 0$ (or $x^{(S)} < 0$, $y^{(S)} > 0$, $y^{(S)} < 0$) region reflects and propagates to the $x^{(S)} < 0$ (or $x^{(S)} > 0$, $y^{(S)} < 0$, $y^{(S)} > 0$) region. The experimental beam profiles obtained from the setup of Fig. \ref{Fig_ExpSetup} are thus $180^\circ$ rotated about $\hat{\mathbf{z}}$, as compared to the simulated profiles obtained from the system of Fig. \ref{Fig_System}. However, the field functions [Eqs. (\ref{ExyXY_comps}), (\ref{terms_def})] are invariant under the transformation $(x,y) \rightarrow (-x,-y)$; because of which, all simulated profiles are symmetric under a $180^\circ$ rotation about $\hat{\mathbf{z}}$ [Figs. \ref{Fig_ExyXY_Profiles}--\ref{Fig_0PM_IE_Profiles}]. For this reason, explicit consideration of such rotations of the experimentally obtained profiles is not required. This observation, nevertheless, reveals another subtle comparison between the diverging and converging beam models.

\subsection{Results}

\begin{figure*}
\begin{center}
\includegraphics[width = 0.75\linewidth]{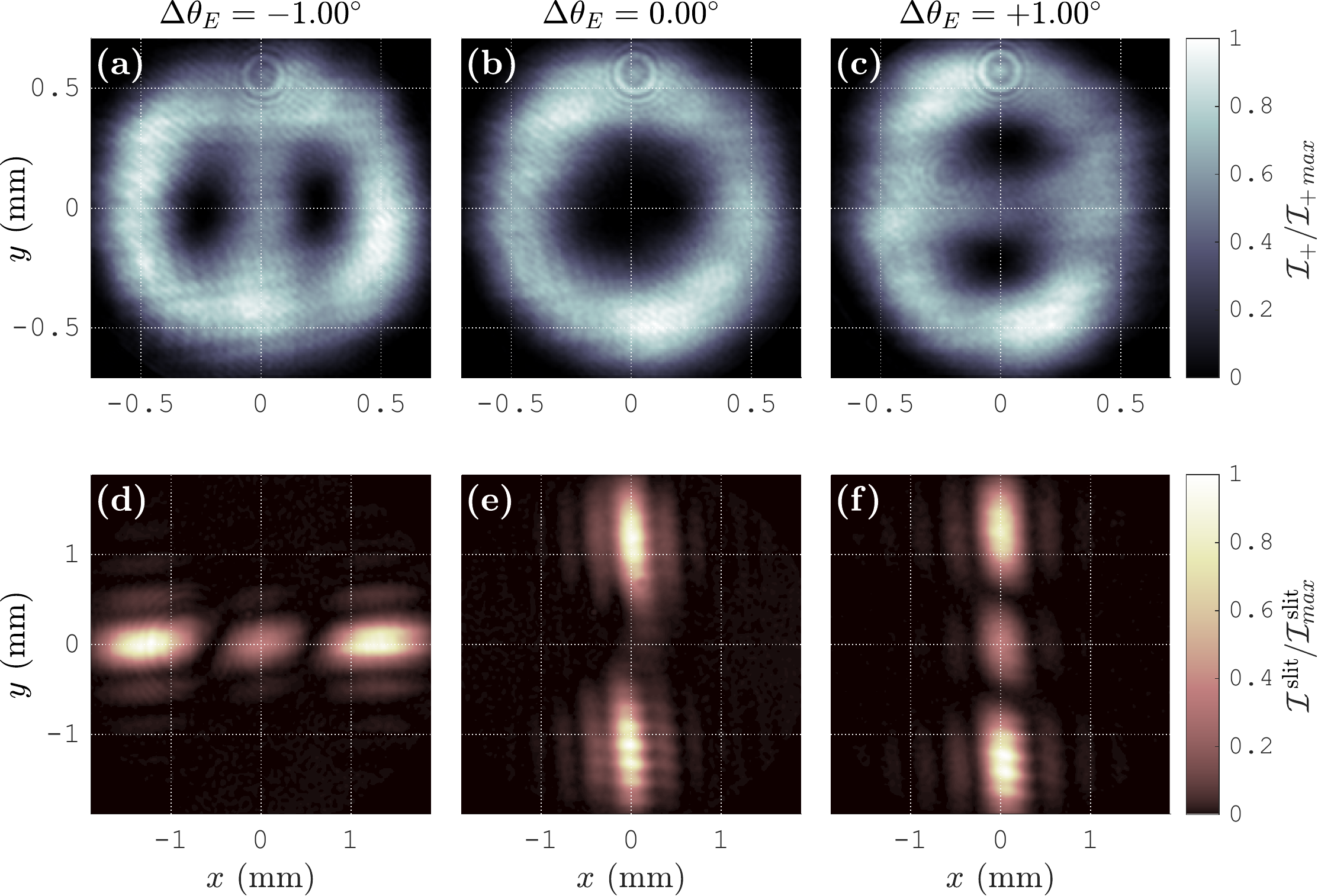}
\end{center}
\caption{\textbf{(a), (b), (c)} Experimentally obtained $\mathcal{I}_+$ intensity profiles for $\Delta\theta_E = -1.00^\circ, 0.00^\circ, +1.00^\circ$ and $\Phi_E = +\pi/2$; \textbf{(d), (e), (f)} Corresponding single-slit diffraction patterns. The diffraction patterns exhibit appropriate fringe dislocations at the positions of the singularities (the length scales of (d), (e), (f) are different from those of (a), (b), (c), because the beam is expanded before passing through the single-slit).
In (e), as we move along a central bright fringe towards the singularity, the fringe bends left and aligns with the first order bright fringe on the other side of the singularity. This identifies a topological charge $t = -2$. In (d) and (f), as we move in a similar way, the central bright fringe bends left and aligns with the first dark fringe on the other side, and also partially connects to the first order bright fringe. This identifies a topological charge $t = -1$. 
\textbf{A Technical Aspect:} In any directly displayed beam image, the $y$ pixel count starts at the top and proceeds downwards. But in a formal profile plot on the $xy$ plane, the $\hat{\mathbf{y}}$ direction is vertically upwards. So the direct image is flipped vertically with respect to the true profile. The images of the single-slit diffraction patterns shown in Ref. \cite{SingleSlit} are thus flipped. The true patterns for $t = \pm1$ are shown in Ref. \cite{BekshaevRev2011}, and are explained in terms of azimuthal energy flows in vortex beams.
}
{\color{grey}{\rule{\linewidth}{1pt}}}
\label{Fig_ExpProfiles}
\end{figure*}

We first impose the central singularity conditions $(\theta_E, \Phi_E) = (45^\circ, +\pi/2)$ by appropriately orienting $G_1$ and $Q_1$ 
\footnote{\textbf{Note:} The laser from the source is linearly polarized. The role of $H$ is to rotate its polarization direction, so that the intensity after $G_1$ is maximized.}. We then rotate $G_1$ to get different off-central angles $\theta_E = 45^\circ + \Delta\theta_E$, and observe the corresponding $\mathcal{I}_+$ intensity profiles. As anticipated, a central singularity is observed for $\Delta\theta_E = 0.00^\circ$; a pair of symmetrically off-centered singularities on the $x$ axis is observed for $\Delta\theta_E < 0.00^\circ$; and a pair of symmetrically off-centered singularities on the $y$ axis is observed for $\Delta\theta_E > 0.00^\circ$. As $\Delta\theta_E$ approaches zero from a non-zero value, the off-central singularities gradually approach towards each other and eventually merge at the center to form the central singularity. 
Example $\mathcal{I}_+$ profiles, for $\Delta\theta_E = -1.00^\circ,0.00^\circ,+1.00^\circ$, are shown respectively in Figs. \ref{Fig_ExpProfiles}(a), \ref{Fig_ExpProfiles}(b) and \ref{Fig_ExpProfiles}(c) --- 
which exhibit behaviors similar to those of
the simulated profiles of Figs. \ref{Fig_0PM_IE_Profiles}(a), \ref{Fig_0PM_IE_Profiles}(b) and \ref{Fig_0PM_IE_Profiles}(c) respectively. This observation clearly verifies the correctness of the simulated singularity dynamics.

Subsequently, we observe the topological charges of these singularities by using single-slit diffraction \cite{SingleSlit, BekshaevRev2011}.
We place a lens in front of $G_S$ to expand the beam size; and then pass the expanded beam through a single slit of width $0.6$ mm, oriented appropriately for the different $\Delta\theta_E$ values (horizontal for $\Delta\theta_E = -1.00^\circ$, vertical for $\Delta\theta_E = 0.00^\circ$ and $+1.00^\circ$). The far-field diffraction patterns thus obtained are shown in Figs. \ref{Fig_ExpProfiles}(d), \ref{Fig_ExpProfiles}(e) and \ref{Fig_ExpProfiles}(f) respectively. In each of Figs. \ref{Fig_ExpProfiles}(d) and \ref{Fig_ExpProfiles}(f), two fringe dislocations appear at the positions of the singularities. Each of these dislocations carry the specific signatures of a topological charge $t = -1$.
This verifies that, for these two fields, the phase profile $\Phi_+$ has a topological charge $t = -1$ corresponding to each of its two singularities, making the total topological charge $t = -2$ [Figs. \ref{Fig_0PM_abpPhp_Profiles}(c), \ref{Fig_0PM_abpPhp_Profiles}(f)]. On the other hand, in Fig. \ref{Fig_ExpProfiles}(e), a single fringe dislocation appears at the position of the central singularity. This dislocation carries the specific signatures of a topological charge $t = -2$; which verifies that, for this field, the phase profile $\Phi_+$ has a single central singularity of topological charge $t = -2$ [Fig. \ref{Fig_abpmPhp_Profiles}(d)].

\begin{figure}
\begin{center}
\includegraphics[width = 0.8\linewidth]{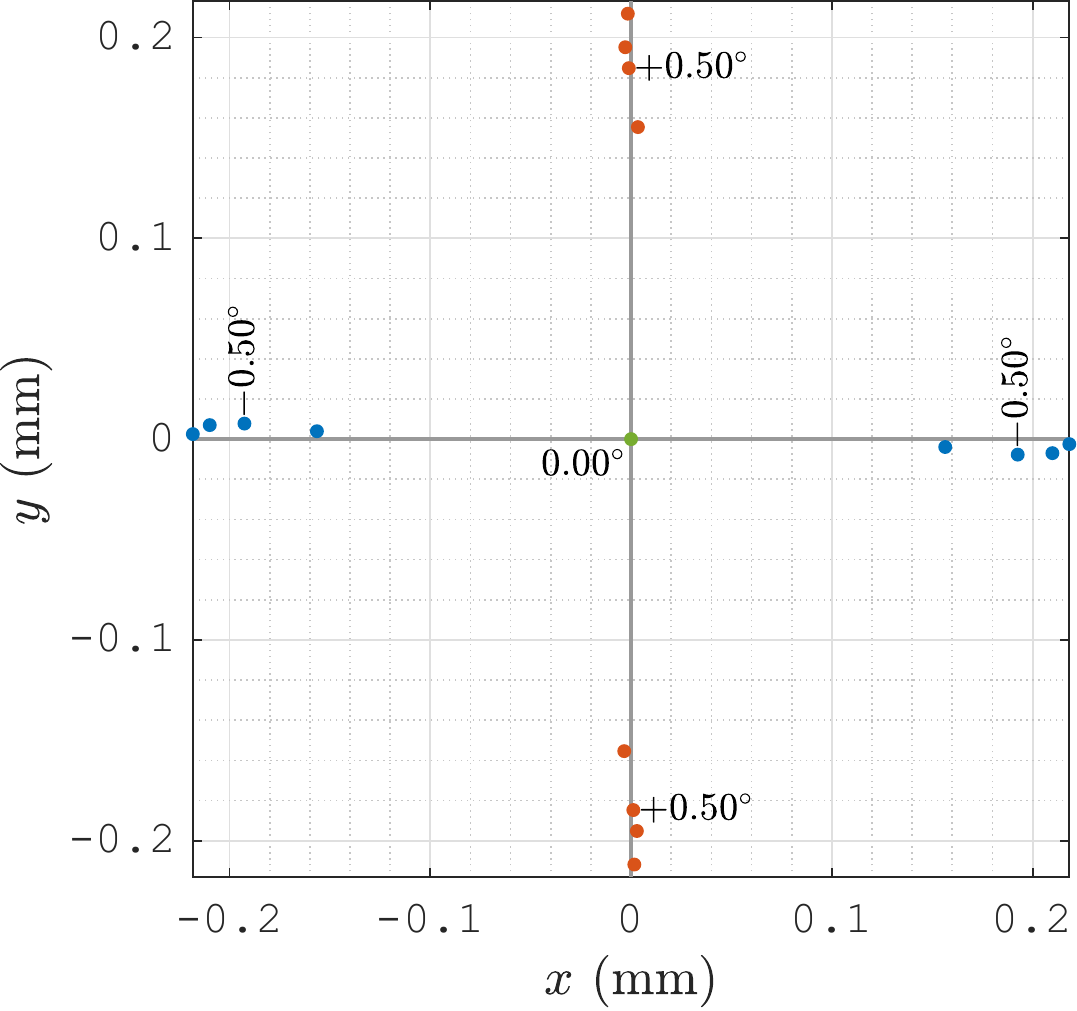}
\end{center}
\caption{Experimentally obtained representative points on the singularity trajectories, with $\Delta\theta_E$ values indicated. Any two adjacent points are separated here by a $\Delta\theta_E$ interval of $0.25^\circ$.}
{\color{grey}{\rule{\linewidth}{1pt}}} \vspace{-1em}
\label{Fig_ExpTrajectory}
\end{figure}

Finally, we capture a series of $\mathcal{I}_+$ profile images for various $\Delta\theta_E$ values, and then computationally determine their singularity coordinates. This generates a series of sample points on the trajectories of the singularities, which is shown in Fig. \ref{Fig_ExpTrajectory}. The distribution pattern of these trajectory points match well with that shown in Fig. \ref{Fig_SingTrajectory}, thus verifying the correctness of the simulated trajectory characteristics.

In this way, the simulated singularity dynamics are experimentally verified. The fact that the diverging and converging beam models create similar dynamics of the singularities is a remarkable result, and is consistent with our assertion [Section \ref{Subsec_SimContext}] that the constituent wavevectors in both these models occupy equivalent regions in momentum space.



\section{Orbital Angular Momentum Characteristics} \label{Sec_SOI}


Since the $\mathcal{E}_+$ field function for $\Delta\theta_E = 0^\circ$ is a pure LG mode of index $l = -2$ [Section \ref{Subsec_CentralSingularity}], the associated OAM is straightforwardly determined as $-2\hbar$ per photon. Such a direct determination, however, cannot be performed for the $\Delta\theta_E \neq 0^\circ$ cases, since the associated phase vortices are non-canonical [Section \ref{Subsec_OffCentralSingularity}].
So, in the present section, we explore the OAM characteristics in detail by explicitly calculating the OAM fluxes associated to the $\boldsymbol{\mathcal{E}}_\pm$ fields.

Barnett's AM flux density formalism \cite{Barnett2002, Gbur} shows that, for any 2D beam field in the form $\bs{\mathcal{E}} = \mathcal{E}_x \, \hat{\mathbf{x}} + \mathcal{E}_y \, \hat{\mathbf{y}}$ (considering an isotropic dielectric medium of refractive index $n$), the OAM flux density across a beam cross-section is given by
\begin{equation}
M_{orb} = \dfrac{n \epsilon_0}{2k} \mathfrak{Im} \left[ (\partial_\phi \mathcal{E}_x) \mathcal{E}_x^* + (\partial_\phi \mathcal{E}_y) \mathcal{E}_y^* \right]; \label{Morb_def}
\end{equation}
where, $k$ is the wavevector magnitude $2\pi/\lambda$ in free space; $\mathfrak{Im}(Z)$ represents the imaginary part of a complex quantity $Z$; and $\partial_\phi = x\partial_y - y\partial_x$ is the partial differential operator with respect to the azimuthal variable $\phi$. Using this formula for the $\boldsymbol{\mathcal{E}}_\pm$ fields in the general form $\mathcal{E}_\pm \, \hat{\boldsymbol{\sigma}}^\pm$, we obtain the general expressions of their OAM flux densities as
\begin{equation}
M_{orb}^\pm = \dfrac{n \epsilon_0}{2 k} \mathfrak{Im} [\mathcal{E}_\pm^* (\partial_\phi \mathcal{E}_\pm)]. \label{MorbPM_def}
\end{equation}
For the presently considered fields, we use Eqs. (\ref{ExyXY_comps})--(\ref{sigmaPM_fields}) in Eq. (\ref{MorbPM_def}) along with $\Phi_E = +\pi/2$, and obtain
\begin{eqnarray}
&& M_{orb}^\pm = -\dfrac{n_1 \epsilon_0}{2 k} \dfrac{C_1^2 C_2}{2} \left[ C_2 s_E c_E \sigma^4 \right. \nonumber\\ 
&& \hspace{1.2em} \left. \pm \left\{ \left( c_E^2 - s_E^2 \right) \left(v^2 - u^2\right) + C_2 \sigma^2 \left( c_E^2 u^2 + s_E^2 v^2 \right) \right\} \right] \! . \hspace{1.6em}
\label{MorbPM_full}
\end{eqnarray}
The corresponding OAM fluxes across the entire beam cross-section is then obtained as
\begin{equation}
L_{orb}^\pm \! = \!\! \int_{-\infty}^\infty \! \int_{-\infty}^\infty \!\! M_{orb}^\pm \, dx \, dy. \label{LorbPM_def}
\end{equation}

\begin{figure*}
\begin{center}
\includegraphics[width = 0.88\linewidth]{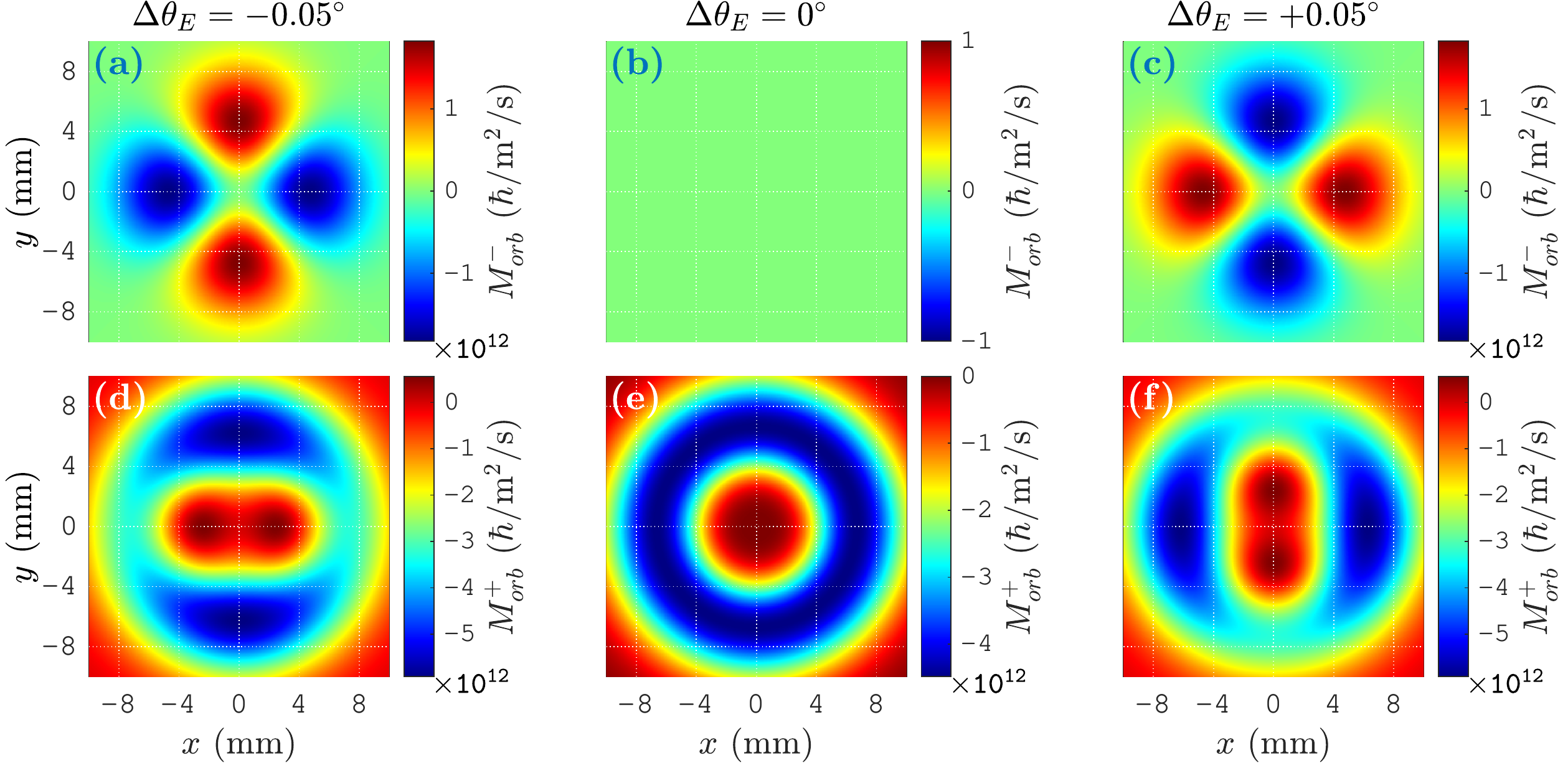}
\end{center}
\caption{OAM flux density profiles $M_{orb}^\pm$ [Eq. (\ref{MorbPM_full})] of the fields $\boldsymbol{\mathcal{E}}_\pm$ [Eq. (\ref{E=Ep+Em})]: \textbf{(a), (b), (c)} $M_{orb}^-$ for $\Delta\theta_E = -0.05^\circ, 0^\circ, +0.05^\circ$; \textbf{(d), (e), (f)} $M_{orb}^+$ for $\Delta\theta_E = -0.05^\circ, 0^\circ, +0.05^\circ$.}
{\color{grey}{\rule{\linewidth}{1pt}}} \vspace{-0.5em}
\label{Fig_MorbProfiles}
\end{figure*}

Example $M_{orb}^\pm$ profiles for $\Delta\theta_E = -0.05^\circ, 0^\circ, +0.05^\circ$ (corresponding to the $\boldsymbol{\mathcal{E}}$ profiles of Figs. \ref{Fig_0PM_IE_Profiles}(d), \ref{Fig_0PM_IE_Profiles}(e), \ref{Fig_0PM_IE_Profiles}(f) respectively) are shown in Fig. \ref{Fig_MorbProfiles}.
As understood from Barnett's analysis \cite{Barnett2002, Gbur}, the fluxes $L_{orb}^\pm$ [Eq. (\ref{LorbPM_def})] are physically significant quantities; but a straightforward physical significance of the flux densities $M_{orb}^\pm$ [Eqs. (\ref{MorbPM_def}), (\ref{MorbPM_full}), Fig. \ref{Fig_MorbProfiles}] is restricted 
(for instance, in Figs. \ref{Fig_MorbProfiles}(d) and \ref{Fig_MorbProfiles}(f), the local positive $M_{orb}^+$ values in the vortex neighborhoods are apparently inconsistent with the topological charge $t = -1$ of each vortex). 
The restriction applies because, any function $M$ that satisfies
\begin{equation}
\int_{-\infty}^\infty \! \int_{-\infty}^\infty \!\! M \, dx \, dy = 0 \label{Mfunction0}
\end{equation}
can be seamlessly added to $M_{orb}^\pm$ without changing the physically significant $L_{orb}^\pm$ results of Eq. (\ref{LorbPM_def}). Such a function $M$ can be chosen, for example, by requiring to compensate for apparent inconsistencies posed by the local $M_{orb}^\pm$ values.
Nevertheless, a certain degree of relevance of the $M_{orb}^+$ profiles of Figs. \ref{Fig_MorbProfiles}(d), \ref{Fig_MorbProfiles}(e) and \ref{Fig_MorbProfiles}(f) can be clearly observed with the $\Phi_+$ phase profiles of Figs. \ref{Fig_0PM_abpPhp_Profiles}(c), \ref{Fig_abpmPhp_Profiles}(d), \ref{Fig_0PM_abpPhp_Profiles}(f) and with the $\mathcal{I}_+$ intensity profiles of Figs. \ref{Fig_0PM_IE_Profiles}(a), \ref{Fig_0PM_IE_Profiles}(b), \ref{Fig_0PM_IE_Profiles}(c).
As interpreted by Berry \cite{Berry1998} and Molina-Terriza et al. \cite{MolinaTerriza2001}, the OAM of a vortex beam is carried by the field surrounding the vortex, but not by the vortex itself which is devoid of photons. By comparing the present $M_{orb}^+$, $\Phi_+$ and $\mathcal{I}_+$ profiles, we see that the regions with larger intensities possess larger magnitudes of $M_{orb}^+$, as compared to the less intense regions in the neighborhood of the vortices. This observation is indeed in agreement with the interpretation by Berry and Molina-Terriza et al.

\begin{figure*}
\begin{center}
\includegraphics[width = 0.95\linewidth]{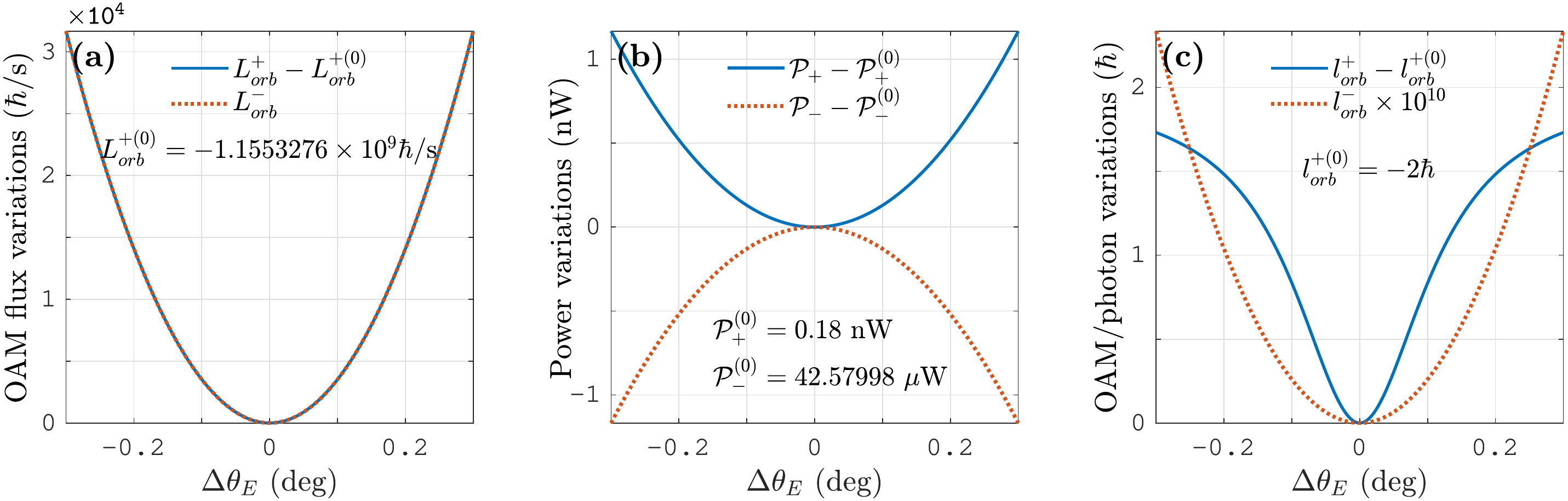}
\end{center}
\caption{\textbf{(a)} Variations of the OAM fluxes $L_{orb}^\pm$ [Eq. (\ref{LorbPM_def})] as functions of $\Delta\theta_E$. Both the functions $L_{orb}^\pm$ have identical variations around the central values (at $\Delta\theta_E = 0^\circ$) $L_{orb}^{+(0)}$ and $0 \hbar/s$.
\textbf{(b)} Variations of the powers $\mathcal{P}_\pm$ [Eq. (\ref{Ppm_def})] as functions of $\Delta\theta_E$. The functions $\mathcal{P}_\pm$ have exact opposite variations around the central values $\mathcal{P}_{\pm}^{(0)}$ --- implying some energy transfer from the field $\boldsymbol{\mathcal{E}}_-$ to the field $\boldsymbol{\mathcal{E}}_+$ as $|\Delta\theta_E|$ moves away from $0^\circ$, while satisfying energy conservation. This energy gain of $\boldsymbol{\mathcal{E}}_+$ is consistent with the evolution of the $\mathcal{I}_+$ profiles of Fig. \ref{Fig_0PM_IE_Profiles}. 
\textbf{(c)} Variations of the `OAM per photon' $l_{orb}^\pm$ [Eq. (\ref{lorbPM})] as functions of $\Delta\theta_E$. The functions $l_{orb}^\pm$ vary around the central values $-2\hbar$ and $0 \hbar$. The variation of $l_{orb}^-$ is represented here by multiplying it with a factor $10^{10}$.
}
{\color{grey}{\rule{\linewidth}{1pt}}} \vspace{-1em}
\label{Fig_LPwlorb}
\end{figure*}

The OAM fluxes $L_{orb}^\pm$ physically signify the rate of flow of OAM across the entire beam cross-section due to the individual $\hat{\boldsymbol{\sigma}}^\pm$ spin-polarized fields. It is very difficult to analytically evaluate the integrals of Eq. (\ref{LorbPM_def}) by using the $M_{orb}^\pm$ expressions of Eq. (\ref{MorbPM_full}); and it is sufficient to perform numerical integration here for the purpose of the present paper.
The numerically determined variations of $L_{orb}^\pm$ for the presently considered $\boldsymbol{\mathcal{E}}_\pm$ fields, in an example $\Delta\theta_E$ range $[-0.3^\circ,+0.3^\circ]$, is shown in Fig. \ref{Fig_LPwlorb}(a). Clearly, $L_{orb}^+$ is several orders of magnitudes higher than $L_{orb}^-$, which is a direct manifestation of the fact that the phase profile $\Phi_+$ has complicated spatial variations (including singularities) at the beam cross-section, whereas the phase $\Phi_- = \tan^{-1} (b_-/a_-)$ is approximately constant (expressed by using Eqs. (\ref{ExyXY_comps})--(\ref{sigmaPM_fields})).

But additionally, the power comparison of the fields $\boldsymbol{\mathcal{E}}_\pm$ makes this difference even more significant. We calculate the powers $\mathcal{P}_\pm$ by integrating the intensities $\mathcal{I}_\pm = (n_1/2\mu_0 c) |\boldsymbol{\mathcal{E}}_\pm|^2$ as
\begin{equation}
\mathcal{P}_\pm = \int_{-\infty}^\infty \! \int_{-\infty}^\infty \!\! \mathcal{I}_\pm \, dx \, dy. \label{Ppm_def}
\end{equation}
The numerically obtained variations of $\mathcal{P}_\pm$ in the $\Delta\theta_E$ range $[-0.3^\circ,+0.3^\circ]$ are shown in Fig. \ref{Fig_LPwlorb}(b) --- clearly expressing the significant dominance of $\mathcal{P}_-$ over $\mathcal{P}_+$. This implies that, as compared to the field $\boldsymbol{\mathcal{E}}_-$, the field $\boldsymbol{\mathcal{E}}_+$ is able to transmit a much higher amount of OAM across a beam cross-section [Fig. \ref{Fig_LPwlorb}(a)] in spite of having a much smaller number of photons [Fig. \ref{Fig_LPwlorb}(b)]. Hence, the quantity `OAM per photon' of the field $\boldsymbol{\mathcal{E}}_+$ massively outweighs that of the field $\boldsymbol{\mathcal{E}}_-$.

Since each photon has an energy $\hbar \omega$, the numbers of $\hat{\boldsymbol{\sigma}}^\pm$ spin-polarized photons passing through a beam cross-section per unit time are obtained as $N_\pm = \mathcal{P}_\pm/\hbar \omega$. The corresponding `OAM per photon' are thus obtained as 
\begin{equation}
l_{orb}^\pm = (L_{orb}^\pm/\mathcal{P}_\pm)\hbar\omega. \label{lorbPM}
\end{equation}
The numerically obtained variations of $l_{orb}^\pm$ in the $\Delta\theta_E$ range $[-0.3^\circ,+0.3^\circ]$ are shown in Fig. \ref{Fig_LPwlorb}(c). Clearly, $l_{orb}^-$ is practically insignificant as compared to $l_{orb}^+$. The physical significance of the $l_{orb}^+$ variation can be understood in reference to the variation of the $\Phi_+$ profile. As explained earlier via Eqs. (\ref{abp_specialCase}) and (\ref{PhiPlus_def}), $\Phi_+$ attains the form $\Phi_+ = \pi - 2\phi$ for $\Delta\theta_E = 0^\circ$ [Fig. \ref{Fig_abpmPhp_Profiles}(d)], because of which the field function $\mathcal{E}_+$ [Eq. (\ref{E=Ep+Em})] becomes a pure LG mode with index $l = -2$. So, with physical consistency, the corresponding OAM per photon attains the value $l_{orb}^+ = -2\hbar$, as understood from the plot of Fig. \ref{Fig_LPwlorb}(c). With the variation of $\Delta\theta_E$, as the function $\mathcal{E}_+$ distorts away from a pure LG mode, complicated $\Phi_+$ profiles such as those of Figs. \ref{Fig_0PM_abpPhp_Profiles}(c) and  \ref{Fig_0PM_abpPhp_Profiles}(f) are obtained. The vortices in these phase profiles are non-canonical \cite{Berry1998, OAMBook2013}; and hence, even if a pair of vortices with topological charges $t = -1$ is obtained, the total OAM per photon does not add up to $-2\hbar$. This characteristic is clearly observed in the plot of Fig. \ref{Fig_LPwlorb}(c).

With the above understanding of the OAM properties, a subtle interpretation and mathematical characterization of SOI can be attained. We re-express the field $\boldsymbol{\mathcal{E}}$ [Eq. (\ref{E=Ep+Em})] as
\begin{eqnarray}
& \boldsymbol{\mathcal{E}} = \mathcal{E}_{0+} e^{i\Phi_+} \hat{\boldsymbol{\sigma}}^+ + \mathcal{E}_{0-} e^{i\Phi_-} \hat{\boldsymbol{\sigma}}^-, \hspace{1em} \mathcal{E}_{0\pm} = \left( a_\pm^2 + b_\pm^2 \right)^\frac{1}{2} \! ; \hspace{1.5em} & \label{E_NonSeparableState}
\end{eqnarray}
which is a non-separable state representation, describing the coupling between the spatial (associated to OAM) and polarization (associated to spin) states. In terms of a semiclassical single-photon interpretation, this state implies that, if a single photon is identified to be in the spin state $\hat{\boldsymbol{\sigma}}^\pm$, it is also automatically identified to be in the spatial state $\mathcal{E}_{0+} e^{i\Phi_\pm}$, which is associated to an OAM $l_{orb}^\pm$ per photon as determined by Eq. (\ref{lorbPM}). This coupled state interpretation reveals an inherent subtle nature of SOI in the considered system. 
The SOI truly takes place due to the inhomogeneous reflection process of the composite beam (i.e., due to the field transformations discussed in the Appendix), because of which the SAM and OAM get coupled. But due to the conservation of AM, the information of this SOI is retained in the subsequent collimated beam. The above non-separable state representation signifies this inherent SOI information. One may qualitatively visualize that, even if an observation is not made on the direct reflected beam, the SOI signature is observed in the subsequent collimated beam in terms of the above non-separable state.
In the classical field picture, this field state implies that the total OAM of the beam field is consistently distributed among the $\hat{\boldsymbol{\sigma}}^\pm$ spin-component fields --- which is a fundamentally interesting physical phenomenon in the presently considered system.


\section{Future Directions} \label{Sec_NearNormal}


Due to the special symmetries of the field terms $\mathcal{E}_x^X$, $\mathcal{E}_y^X$, $\mathcal{E}_x^Y$ and $\mathcal{E}_y^Y$ in the normal reflection [Eqs. (\ref{ExyXY_comps}), (\ref{terms_def}); Fig. \ref{Fig_ExyXY_Profiles}], we have obtained the centroids of the intensity profiles $\mathcal{I}$ and $\mathcal{I}_\pm$ of the fields $\boldsymbol{\mathcal{E}}$ and $\boldsymbol{\mathcal{E}}_\pm$ at the origin --- implying zero GH, IF and spin shifts. But we have observed in the simulation that, as we move away from the normal incidence condition, the symmetries of the field terms break down. The appearance of a unique central plane of incidence breaks the degeneracy of the normal incidence case; and hence non-zero beam shifts and spin shifts appear.

A detailed analysis of a near-normal incidence case cannot be performed by using a usual first order approximation as in Ref. \cite{BARev}, because in such a case the angle of incidence is smaller than or comparable in size to the divergence angle of the beam. Special analyses on near-normal incidence cases are present in the literature \cite{ZhuPBPhaseSHE2021, MazanovAnomalous2021}; but, to our knowledge, without a complete mathematical characterization of the OAM fluxes.
Such an analysis is out of the scope of the present discussion under the mathematical construct of the reflection and transmission coefficient matrices \cite{ADNKVrt2020}. For a future analysis in this direction, we anticipate the appearance of fundamentally interesting phenomena regarding dynamics of singularities and evolution of SOI characteristics as the system transits from a normal incidence to a near-normal incidence configuration. Looking further forward, such an analysis would be substantially significant for dielectric interfaces involving anisotropic, gyrotropic and topological materials; and also for ellipsoidal, helical and other non-planar incident wavefronts.


Additionally, in the present work we have experimentally demonstrated the singularity dynamics, but not the OAM flux determination. To our knowledge, an experimental method utilizing Barnett's formalism to determine OAM flux is not present in the literature; and introducing a new method in this regard is out of the scope of the present paper. We anticipate that such a future method would be a general one, applicable to a large class of beam fields. Our work in this direction, with an appropriate methodology and its experimental demonstration, will be reported elsewhere.



\vspace{-0.5em}

\section{Conclusion} \label{Sec_Conc}

In this paper we have analyzed the degenerate case of normal incidence and reflection of a composite optical beam, which is characterized by an azimuthally symmetric variation of Fresnel coefficients with respect to wavevector orientations, and an inhomogeneously polarized incident beam field. We have used the reflection and transmission coefficient matrix formalism to derive an exact expression of the reflected beam field --- thus completely characterizing its fundamental polarization inhomogeneity.
Then we have introduced appropriate conditions to generate a second order phase singularity in the $\hat{\boldsymbol{\sigma}}^+$ spin-polarized component field, and an associated `center' polarization-singular pattern in the total field. Subsequently, we have introduced controlled variations to the system to deviate from the central singularity generation condition, leading to a splitting of the higher order central singularities (in both phase and polarization) to off-central pairs of lower order singularities. We have thus observed a complex dynamics of the singularities, along with appropriately mapping the trajectories of the split lower order singularities.

Subsequently, we have experimentally observed the singularity dynamics by using a standard focused beam setup for normal incidence. In particular, we have demonstrated the formation and splitting of a higher order singularity, and the trajectories of the split pair of lower order singularities.
The confirmation that the observed intensity minima are truly the anticipated optical vortices has been achieved via single-slit diffraction patterns.

Finally, we have mathematically characterized the SOI information in the system by determining the associated OAM fluxes via Barnett's formalism.
Our simulation has shown the restructuring of the OAM flux densities due to the displacements of the split pair of phase singularities along the determined trajectories.
The $\hat{\boldsymbol{\sigma}}^\pm$ spin-polarized component fields carry their individual OAM fluxes, implying a consistent distribution of OAM between the spin-component fields. At a single-photon level, this implies that a photon detected in a spin state $\hat{\boldsymbol{\sigma}}^\pm$ has a specific OAM $l_{orb}^\pm$ [Eq. (\ref{lorbPM})]. This interpretation reveals a subtle and interesting signature of SOI in the considered system.

In this way, we have explored a significant optical singularity dynamics and analyzed the SOI phenomena in the considered normal-reflected beam field. 
In addition to the practical importance of the normal incidence analysis, the present work also demonstrates a significant application of Barnett's AM flux density formalism. The transfer of OAM from one medium to another is naturally determined by the OAM flux; and hence the involvement of Barnett's formalism is physically relevant and significant. However, to our knowledge, the true potential of this formalism has not been widely explored in the literature. In this premise, our complete characterization of OAM fluxes serves the purpose of demonstrating the true potential of Barnett's formalism.
We anticipate that the exact information on the OAM fluxes, in addition to the controlled variation of the singularity structures, will find significant applications in interface characterization, particle trapping and rotation, and various other nano-optical processes.

\section*{A\lowercase{ppendix}: D\lowercase{erivation of} E\lowercase{qs.} (\ref{E(R)_final})--(\ref{terms_def})} \label{App_DeriveE}

Here we summarize the calculation steps required for deriving the final output field expression [Eqs. (\ref{E(R)_final})--(\ref{terms_def})]. We apply the formalism of Ref. \cite{ADNKVrt2020} to the system of Fig. \ref{Fig_System} for this purpose. Since the considered initial and final beams are collimated, it is sufficient to suppress all $\mathbf{k}\cdot\mathbf{r} - \omega t$ phase terms, and discuss the transformations of the field amplitude vectors only.

From the system geometry it is obtained that, if the 2D coordinates of the point $P_I$ (and hence of $P_0$) are $(x^{(I)},y^{(I)}) = (x_I,y_I)$, then those of the point $P_R$ (and hence of $P$) are $(x,y) = (-\alpha x_I, \alpha y_I)$, where $\alpha = \mathcal{F}_2/|\mathcal{F}_1|$. The input field before $L_1$ at $(x^{(I)},y^{(I)}) = (x_I,y_I)$ thus gives the output field at $(x,y) = (-\alpha x_I, \alpha y_I)$ at $S_R$ after appropriate transformations.

The field transformation at $P_I$ (from `just before $L_1$' to `just after $L_1$', in terms of the $I$ coordinate system) is given by the operator
\begin{equation}
\tilde{\mathbf{R}}_{L1} = g_I\,  \tilde{\mathbf{R}}_{II'} \tilde{\mathbf{R}}_{I'I''} \tilde{\mathbf{R}}_{I'I} \, ; \label{RL1_def}
\end{equation}
\begin{subequations}
\label{RL1_subeqs}
\begin{eqnarray}
& \tilde{\mathbf{R}}_{I'I} = 
\begin{bmatrix}
\cos\phi_I & \sin\phi_I & 0 \\
-\sin\phi_I & \cos\phi_I & 0 \\
0 & 0 & 1
\end{bmatrix} \hspace{-0.2em} ,
\hspace{0.4em} \tilde{\mathbf{R}}_{II'} = \tilde{\mathbf{R}}_{I'I}^{-1} \, ,\hspace{1em}& \label{RIIP} \\
&\tilde{\mathbf{R}}_{I'I''} = 
\begin{bmatrix}
\cos\theta_I & 0 & \sin\theta_I \\
0 & 1 & 0 \\
-\sin\theta_I & 0 & \cos\theta_I
\end{bmatrix} \hspace{-0.2em} ,
\hspace{0.5em} g_I = \dfrac{1}{\sqrt{\cos\theta_I}} \, ; \hspace{1em} 
& \label{RIPPgI}
\end{eqnarray}
\end{subequations}
\begin{subequations}
\label{rI,csthIphI_def}
\begin{eqnarray}
& \mbox{where,} \hspace{1em} \cos\phi_I = x^{(I)}/\rho^{(I)}, \hspace{0.5em}
\sin\phi_I = y^{(I)}/\rho^{(I)}, \hspace{1.5em} & \label{cs_phiI} \\
& \rho^{(I)} = \left( x^{(I)\,2} + y^{(I)\,2} \right)^\frac{1}{2} \!, \hspace{0.5em} r_I = \left( \rho^{(I)\,2} + \mathcal{F}_1^2 \right)^{\frac{1}{2}} \! , \hspace{1.5em} & \label{rI_exact} \\
& \cos\theta_I = |\mathcal{F}_1|/r_I , \hspace{0.5em}
\sin\theta_I = \rho^{(I)}/r_I . \hspace{1.5em} & \label{cs_thetaI}
\end{eqnarray}
\end{subequations}

The field transformation along $P_I \rightarrow P_S \rightarrow P_R$ (from `just after $L_1$' in the $I$ coordinate system, to `just before $L_2$' in the $R$ coordinate system) is given by the operator 
\begin{equation}
\tilde{\mathbf{R}}_{R} = (1/\alpha) \, \tilde{\mathbf{R}}_{RS}\, \tilde{\mathbf{r}}_{S}^{(S)} \tilde{\mathbf{R}}_{SI} \, ; \label{RR_def}
\end{equation}
where $1/\alpha$ is an amplitude reducing factor; $\tilde{\mathbf{R}}_{SI}$ and $\tilde{\mathbf{R}}_{RS}$ are rotation matrices (considering $\theta_{i0} = 0^\circ$)
\begin{equation}
\tilde{\mathbf{R}}_{SI} = 
\begin{bmatrix}
1 & 0 & 0\\
0 & 1 & 0\\
0 & 0 & 1
\end{bmatrix} \! , 
\hspace{1em}
\tilde{\mathbf{R}}_{RS} = 
\left[\begin{array}{ccc}
-1 & 0 & 0\\
0 & 1 & 0\\
0 & 0 & -1
\end{array} \right]
\! ; \label{R_SI,R_RS}
\end{equation}
and $\tilde{\mathbf{r}}_{S}^{(S)}$ is the reflection coefficient matrix
\begin{subequations}
\label{rMat_fulldef}
\begin{eqnarray}
& \tilde{\mathbf{r}}_{S}^{(S)} = \mathcal{A}_0 
\begin{bmatrix}
\mathcal{A}_{11} & \mathcal{A}_{xy} & 0 \\
\mathcal{A}_{xy} & -\mathcal{A}_{10} & 0 \\
0 & 0 & -\mathcal{A}_{01}
\end{bmatrix}; &
\label{rMat}\\
& \mathcal{A}_{pq} = k_{ixS}^{(S)\,2} + (-1)^p\, k_{iyS}^{(S)\,2} + (-1)^q\, k_{tzS}^{(S)}k_{izS}^{(S)} \, , 
\hspace{0.5em} & \label{Apq}\\
& \mathcal{A}_{xy} = 2\, k_{ixS}^{(S)} k_{iyS}^{(S)} \, ,
\hspace{0.5em}
\mathcal{A}_z = \dfrac{k_{tzS}^{(S)} - k_{izS}^{(S)}}{k_{tzS}^{(S)} + k_{izS}^{(S)}}, \hspace{0.5em} \mathcal{A}_0 = \dfrac{\mathcal{A}_z}{\mathcal{A}_{00}} \, ; \hspace{2em} & \label{Axyz0} 
\end{eqnarray}
\end{subequations}
where, $(k_{ixS}^{(S)}, k_{iyS}^{(S)}, k_{izS}^{(S)})$ and $(k_{ixS}^{(S)}, k_{iyS}^{(S)}, k_{tzS}^{(S)})$ are the components of the incident and transmitted local wavevectors $\mathbf{k}_i$ and $\mathbf{k}_t$ at $P_S$ in terms of the $S$ coordinate system; and $p,q = 0,1$.

The field transformation at $P_R$ (from `just before $L_2$' to `just after $L_2$', in terms of the $R$ coordinate system) is an effective inverse of the operation at $P_I$, given by the operator
\begin{equation}
\tilde{\mathbf{R}}_{L2} = (1/g_R)\,  \tilde{\mathbf{R}}_{RR'} \tilde{\mathbf{R}}_{R'R''}^{-1} \tilde{\mathbf{R}}_{R'R} \, ; \label{RL2_def}
\end{equation}
\begin{subequations}
\label{RL2_subeqs}
\begin{eqnarray}
& \tilde{\mathbf{R}}_{R'R} = 
\begin{bmatrix}
\cos\phi_R & \sin\phi_R & 0 \\
-\sin\phi_R & \cos\phi_R & 0 \\
0 & 0 & 1
\end{bmatrix} \hspace{-0.2em} ,
\hspace{0.4em} \tilde{\mathbf{R}}_{RR'} = \tilde{\mathbf{R}}_{R'R}^{-1} \, ,\hspace{1em}& \label{RRRP} \\
&\tilde{\mathbf{R}}_{R'R''} = 
\begin{bmatrix}
\cos\theta_R & 0 & \sin\theta_R \\
0 & 1 & 0 \\
-\sin\theta_R & 0 & \cos\theta_R
\end{bmatrix} \hspace{-0.2em} ,
\hspace{0.5em} g_R = \dfrac{1}{\sqrt{\cos\theta_R}} \, ; \hspace{1em} 
& \label{RRPPgR}
\end{eqnarray}
\end{subequations}
\begin{subequations}
\label{rR,csthRphR_def}
\begin{eqnarray}
& \mbox{where,} \hspace{1em} \cos\phi_R = x/\rho, \hspace{0.5em}
\sin\phi_R = y/\rho, \hspace{1.5em} & \label{cs_phiR} \\
& \hspace{1em} \rho = \left( x^{2} + y^{2} \right)^\frac{1}{2} \!, \hspace{0.5em} r_R = \left( \rho^{2} + \mathcal{F}_2^2 \right)^{\frac{1}{2}} \! , \hspace{1.5em} & \label{rR_exact} \\
& \cos\theta_R = \mathcal{F}_2/r_R , \hspace{0.5em}
\sin\theta_R = \rho/r_R . \hspace{1.5em} & \label{cs_thetaR}
\end{eqnarray}
\end{subequations}

In the initial and final path segments $P_0 \rightarrow P_I$ and $P_R \rightarrow P$, the field amplitude vectors remain unchanged. Then, by compiling the above results, the complete transformation from the initial field $\bs{\mathcal{E}}_0^{(I)}$ (at $P_0$, in terms of the $I$ coordinate system) [Eqs. (\ref{E0I_full})] to the final field $\bs{\mathcal{E}}$ (at $P$, in terms of the $R$ coordinate system) is obtained as
\begin{equation}
\bs{\mathcal{E}} = \tilde{\mathbf{R}}_{L2} \tilde{\mathbf{R}}_{R} \tilde{\mathbf{R}}_{L1} \, \bs{\mathcal{E}}_0^{(I)}. \label{App_E=RRRE0I}
\end{equation}
The system geometry gives the relations
\begin{equation}
\theta_I = \theta_R, \hspace{1em} \cos\phi_I = - \cos\phi_R, \hspace{1em} \sin\phi_I = \sin\phi_R,
\end{equation}
which we use in the expansion and simplification of Eq. (\ref{App_E=RRRE0I}) to obtain the final expressions of Eqs. (\ref{E(R)_final})--(\ref{terms_def}).

Since the choice of the point $P_0$ in the above discussion is arbitrary, the analysis is applicable to all initial points $P_0$ in the input beam field and the corresponding final points $P$ at the screen $S_R$. Hence, Eqs. (\ref{E(R)_final})--(\ref{terms_def}) give the complete field at $S_R$ as a function of $(x,y)$.



\begin{acknowledgments}
A.D. thanks CSIR (India) for Senior Research Fellowship (SRF). N.K.V. thanks SERB (DST, India) for financial support.
\end{acknowledgments}


\bibliography{AD_NKV_NormInc_Refs}

\end{document}